\documentclass[twocolumn,preprintnumbers,amsmath,amssymb,superscriptaddress,nofootinbib,longbibliography]{revtex4-1}
\usepackage{graphicx}
\usepackage{bm}
\usepackage{dsfont}
\usepackage[usenames,dvipsnames]{xcolor}
\usepackage{pstricks}
\usepackage[tight]{subfigure}
\usepackage{verbatim}
\usepackage{units}
\usepackage{multirow}
\usepackage{enumitem}
\usepackage{mathrsfs}
\usepackage{leftidx}
\usepackage{xspace}

\usepackage[a4paper]{hyperref}
\hypersetup{colorlinks=true,linktoc=all,linkcolor=blue,breaklinks=true,citecolor=blue,urlcolor=blue}

\usepackage[english]{babel}
\newcommand{\kB}{k_\text{B}}

\newcommand{\expec}[1]{\langle #1 \rangle}

\newcommand{\op}[1]{\skew{0}{\hat}{#1}}
\newcommand{\opI}{\skew{3}{\hat}{I}}

\newcommand{\intd}{\text{d}}

\newcommand{\etal}{\emph{et al.}\xspace}

\newcommand{\avmu}{\overline{\mu}}

\newcommand{\muL}{\mu_\text{L}}
\newcommand{\TL}{T_\text{L}}

\newcommand{\mup}{\mu_\text{p}}
\newcommand{\avmup}{\overline{\mu}_\text{p}}
\newcommand{\Dmup}{\Delta\mu_\text{p}}
\newcommand{\Tp}{T_\text{p}}
\newcommand{\avTp}{\overline{T}_\text{p}}
\newcommand{\DTp}{\Delta T_\text{p}}

\newcommand{\perT}{\mathcal{T}}
\newcommand{\vol}{\mathcal{V}}

\newcommand{\kappae}{\kappa_{\text{e},0}}
\newcommand{\kappaph}{\kappa_{\text{ph},0}}

\newcommand{\Cq}{C_\text{q}}
\newcommand{\Cg}{C_\text{g}}
\newcommand{\CE}{C_E}

\newcommand{\Is}{I_\text{s}}
\newcommand{\avIs}{\bar{I}_\text{s}}

\newcommand{\avI}{\bar{I}}

\newcommand{\Ip}{I_\text{p}}
\newcommand{\avIp}{\bar{I}_\text{p}}
\newcommand{\DIp}{\Delta I_\text{p}}
\newcommand{\dIp}{\delta I_\text{p}}
\newcommand{\opIp}{\hat{I}_\text{p}}

\newcommand{\Jp}{J_\text{p}}
\newcommand{\avJp}{\bar{J}_\text{p}}
\newcommand{\dJp}{\delta J_\text{p}}
\newcommand{\opJp}{\hat{J}_\text{p}}

\newcommand{\tauE}{\tau_E}
\newcommand{\tauee}{\tau_\text{e-e}}
\newcommand{\tauRC}{\tau_{RC}}
\newcommand{\taueph}{\tau_\text{e-ph}}

\newcommand{\Pp}{\mathcal{P}_\text{p}}
\newcommand{\Po}{\mathcal{P}_0}
\newcommand{\Ps}{\mathcal{P}_\text{s}}
\newcommand{\Psdir}{\mathcal{P}_\text{s,dir}}
\newcommand{\Psind}{\mathcal{P}_\text{s,ind}}

\newcommand{\Fs}{\mathcal{F}_\text{s}}
\newcommand{\Fsdir}{\mathcal{F}_\text{s,dir}}
\newcommand{\Fsind}{\mathcal{F}_\text{s,ind}}

\newcommand{\TR}{T_\text{R}}

\newcommand{\Ds}{D_\text{s}}

\newcommand{\fL}{f_\text{L}}
\newcommand{\fp}{f_\text{p}}

\newcommand{\funH}{\mathcal{H}}

\newcommand{\SF}{S^\text{F}}
\newcommand{\SFcon}{S^{\text{F}\ast}}

\usepackage[normalem]{ulem}

\begin{document}

	
	\title{Probing charge and heat current noise by frequency-dependent temperature\\ and potential fluctuations}
	
	\author{Nastaran Dashti}
	\affiliation{Department of Microtechnology and Nanoscience (MC2), Chalmers University of Technology, S-412 96 G\"oteborg, Sweden}
	
	\author{Maciej Misiorny}
	\affiliation{Department of Microtechnology and Nanoscience (MC2), Chalmers University of Technology, S-412 96 G\"oteborg, Sweden}
	\affiliation{Faculty of Physics, Adam Mickiewicz University, 61-614 Pozna\'{n}, Poland}
	
		\author{Peter Samuelsson}
	\affiliation{Department of Physics and NanoLund, Lund University, S-221 00 Lund, Sweden}
	
	\author{Janine Splettstoesser}
	\affiliation{Department of Microtechnology and Nanoscience (MC2), Chalmers University of Technology, S-412 96 G\"oteborg, Sweden}

	\date{\today}
	
\begin{abstract}
The energetic properties of electron transport in mesoscopic and nanoscale conductors is of large current interest. Here we theoretically investigate the possibility of probing fluctuations of charge and heat currents as well as their mixed correlations via fluctuations of the temperature and electrochemical potential of a probe coupled to the conductor. Our particular interest is devoted to the charge and energy noise stemming from time-dependently driven nanoelectronic systems designed for the controlled emission of single electrons, even though our setup is appropriate for more general AC driving schemes.
We employ a Boltzmann-Langevin approach in order to relate the frequency-dependent electrochemical potential and temperature fluctuations in the probe to the bare charge and energy current fluctuations emitted from the electron source. 
We apply our findings to the prominent example of an on-demand single-electron source, realized by a driven mesoscopic capacitor in the quantum Hall regime. We show that neither the background fluctuations of the probe in the absence of the working source, nor the fluctuations induced by the probe hinder the access to the sought-for direct source noise for a large range of parameters.
\end{abstract}

\maketitle
	
\section{Introduction}
	
Charge transport in mesoscopic and nanoscale structures can nowadays be manipulated at the single-electron level. Of particular interest for different fields, ranging from metrology to quantum information and quantum optics with electrons, are dynamically driven single-electron sources injecting particles into a conductor \textit{on demand}~\cite{Feve2007May,Blumenthal2007May,McNeil2011Sep,Hermelin2011Sep,Roche2013,Dubois2013Oct,Ubbelohde2015Jan}.  A way to access the precision of these sources, which is not accessible by the average charge current alone, is by detecting their charge current statistics. Indeed, already the low-frequency noise (the second moment of the statistics) is known to provide additional information about the number of particles that take part in transport, thereby giving access to electron-hole pair creation detrimental for the source precision~\cite{Vanevic2008Dec,Gabelli2013Feb,Dubois2013Oct}. 

Not only the transported charged particles, but also the neutral electron-hole pair excitations in general carry energy~\cite{Moskalets2009Aug,Rossello2015Mar}. This fact has, in recent years, motivated theoretical investigations on the fluctuations of charge currents, heat currents and also their mixed correlations for driven systems~\cite{Moskalets2014May,Battista2014Aug,Battista2014Dec,Vannucci2017Jun}, see also Refs.~\cite{Zhan2011Nov,Crepieux2014Dec,Sanchez2012Dec} for the stationary bias case. The energetic properties revealed in this way
are of key importance: single-electron sources, while injecting single particles into a conductor in a well-controlled manner, can also be used for quantum state preparation of electronic flying qubits. Therefore, in addition to the precision, knowledge about the energy spectrum of the injected signal is required. 

\begin{figure}[b]
		\centering
		\includegraphics[width=3.in]{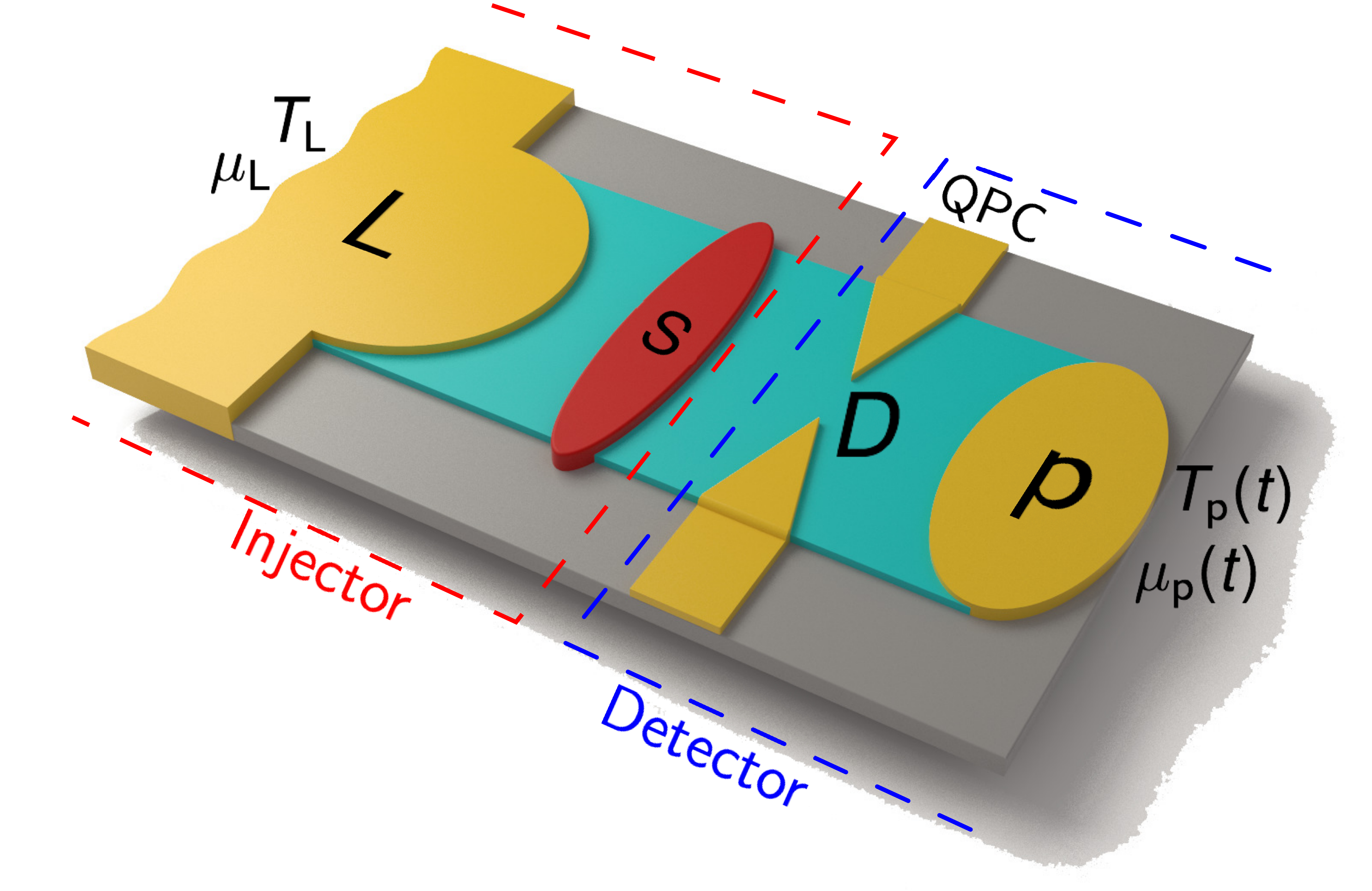}
		\caption{Schematic of the ballistic, two-terminal noise detection system, with the injector and detector parts indicated. The injector part consists of an electronic reservoir, $\text{L}$, kept at an electrochemical potential $\mu_\text{L}$ and temperature $T_\text{L}$, and a time-dependently driven forward scattering potential, S (red). On the detector side, the electrons emitted from the injector are scattered at a QPC with transparency $D$.  The transmitted electrons are absorbed at the electrically and thermally floating probe, p, inducing fluctuations of the electrochemical potential $\mu_\text{p}(t)$ and temperature $T_\text{p}(t)$.}
		\label{fig_setup}
\end{figure} 

An urging question is hence how charge and energy currents as well as their fluctuations can actually be accessed experimentally. Recently, energy-resolved charge currents of single-electron sources have been measured with the help of an energy-dependent detection barrier~\cite{Fletcher2013Nov,Ubbelohde2015Jan}. Even a highly complex state tomography~\cite{Grenier2011Sep,Samuelsson2006Jan} has been performed to access the quantum state of the single particles~\cite{Jullien2014Oct}. In order to detect fluctuations in the heat current, the detection of power fluctuations has been suggested~\cite{Battista2014Aug}, which have a straightforward interpretation when the source is voltage-bias driven.
 	
In this paper, we analyze the possibility of extracting charge current and heat current noise -- injected from an arbitrary AC electronic current source -- as well as their mixed correlations from reading out the temperature- and electrochemical potential fluctuations~\cite{Battista2013Mar,Utsumi2014May,Berg2015Jul} of a floating probe contact~\cite{Buttiker1988Jan,Buttiker1986Oct}, see Fig.~\ref{fig_setup}. We emphasize that for the detection of electrochemical potentials and their fluctuations in a system fed by a steady state charge current, in general conductors with three or more terminals are required~\cite{Buttiker1986Oct}. Instead, for the analysis of the fluctuations of charge and heat currents of a \textit{pure AC current source}, the simple setting suggested here suffices.
Probes of similar type have been used for stationary heat current detection~\cite{Gasparinetti2011May}. Moreover, fast temperature measurement techniques are nowadays available~\cite{Gasparinetti2015Jan} and motivate our theoretical proposal.

Using a Boltzmann-Langevin approach~\cite{Blanter2000Sep}, we directly relate the low-frequency fluctuations of charge and energy currents flowing into the probe to the finite-frequency probe temperature and potential fluctuations. The frequency-dependence of the fluctuations sets a fundamental constraint on the bandwidth within which the fluctuations can be detected. In order to furthermore extract the \textit{bare source} fluctuations, which we are interested in, from the total noise detected with the help of the probe, one has to proceed as follows: first, subtract the \textit{pure probe} fluctuations (obtained when the source is not operating) and, second, disentangle the remaining data from the fluctuations \textit{induced} by the temperature and potential measurements.

As a proof-of-principle, we apply our findings to a case where a time-dependently driven mesoscopic capacitor~\cite{Feve2007May} in the quantum Hall regime is taking the role of the current injector. The noise~\cite{Blanter2000Sep} stemming from the source and serving as the bare fluctuations entering the Boltzmann-Langevin approach, is derived for this explicit example by employing (Floquet) scattering theory~\cite{Moskalets2011Sep}. We can thereby demonstrate the possibility of determining the source fluctuations from the potential and temperature fluctuations of the probe. 

Finally, we want to point out that the relation between current noise and detectable dissipated energy is of much ampler relevance. The relation between higher moments of charge current statistics and power fluctuations~\cite{Battista2013Mar,Tikhonov2014Oct} has, for example, been exploited in the detection of photons emitted from the noisy current through a tunneling barrier~\cite{Gabelli2004Jul,Simoneau2017Feb,Mora2017Mar}. Vice versa, heat currents are presently detected via noise measurements~\cite{Jezouin2013Oct}

This paper is organized as follows: In Sec.~\ref{sec:model} we introduce the setup, consisting of an injector to be analyzed and the detector being made of a probe contact and a quantum point contact (QPC). The theoretical Boltzmann-Langevin approach is introduced in Sec.~\ref{sec:formalism} and employed to the calculation of the probe fluctuations in Sec.~\ref{sec:probefluct}. We then,  in Sec.~\ref{sec:decompose}, separate the fluctuations into pure probe fluctuations and fluctuations arising when the source is switched on, including the sought-for bare source fluctuations. In Sec.~\ref{sec:example} we analyze the specific case of a time-dependently driven mesoscopic capacitor playing the role of the injector. 

\section{Model}\label{sec:model}

In this section, we  introduce the model setup for the generation and, in particular, for the detection of fluctuating charge and energy currents and their correlations. Next, we discuss the relevant timescales of the problem, and the fundamental requirements for an experimental realization of our proposal.

\subsection{Description of the setup}\label{sec:model_setup}

A sketch of the mesoscopic noise detection system, implemented in a ballistic, two-terminal conductor, is presented in Fig.~\ref{fig_setup}. The system comprises two key components referred to as an \emph{injector} and a \emph{detector}.
The \emph{injector} part consists of an electronic reservoir (denoted by~`L'), electrically grounded~(\mbox{$\muL=0$}) and kept at a fixed temperature~$\TL$, and a scattering region (denoted by~`S') representing an AC current source \mbox{-- a}~generic device injecting electrons (and/or holes) into the conductor. 
 Such a current source can be analyzed via its current and noise properties. Our goal is to investigate how the latter can be accessed by measuring fluctuations of the electrochemical potential~$\mup(t)$ and temperature~$\Tp(t)$ in the probe (denoted by~`p') belonging to the detector part of the setup. 
We note that the probe scheme under consideration is applicable for any rapidly fluctuating AC current source. 
In fact, the only important assumptions required at this point are: 
(i)~the lack of back-reflection between the source scattering region and a quantum point contact (QPC) of the detector region, which could otherwise lead to the creation of resonant states at the injector/detector
interface;\footnote{%
	Note that the addition of a floating (but energy-conserving) contact at the outcome of the QPC could be used to suppress emerging resonances, if necessary.
}
(ii) the charge- and energy-current source is characterized by some timescale~$\perT$ associated with the rate of charge (and energy) emission.

An illustrative example of the AC current source that we consider in this paper is a time-dependently driven mesoscopic capacitor injecting on-demand single particles into the conductor. The operation of such a source, experimentally demonstrated by F\`{e}ve \etal~\cite{Feve2007May}, can be theoretically described  by a Floquet scattering matrix approach~\cite{Dashti17} -- for details see Sec.~\ref{sec:example}. 
Importantly, in the case of such a periodically driven current source,  the natural characteristic timescale associated with the source is the driving period~$\perT$, related to the driving frequency as~\mbox{$\Omega=2\pi/\perT$}. For the energetic properties of the source, also the width in time of the injected current pulse, $\sigma$, will turn out to be important, see Sec.~\ref{sec:example}.

On the other hand, the \emph{detector} part of the setup contains a QPC with a single active transport mode, and a thermally floating reservoir acting as a probe, indicated by `p' in Fig.~\ref{fig_setup}. 
The electrical conductance of the QPC is~$gD$, where $D$ denotes the tunable, energy-independent transparency and \mbox{$g=e^2/h$} is the conductance quantum per spin, with the electron charge $-e$ (that is, \mbox{$e>0$}). Having in mind an injector system which operates in the fully spin-polarized quantum Hall regime, we discard here  the electronic spin-degree of freedom. 
At the QPC, the fluctuations of the injector currents are modified due to elastic scattering. This modification, depending on $D$, is used to tune the different components of the probe noise, as further discussed in Sec.~\ref{sec:meso_fluct}. 
The core element of the detector is, however, the \textit{floating probe}. We assume it to be metallic and big enough to avoid Coulomb-blockade effects. The probe works in the hot-electron regime, meaning that particles entering the probe thermalize quickly via electron-electron scattering. For this reason, a well-defined electrochemical potential~$\mup(t)$ and electronic temperature~$\Tp(t)$ can be associated with the probe at all times~$t$. 
Moreover, we assume that the statistics of the energy dissipated in the probe is determined by the energy statistics of the signal entering the
probe.\footnote{%
	Interestingly, the validity of this assumption can actually be tested with the setup suggested here, by applying a driving signal with known statistics and detecting the resulting probe fluctuations.
}

Importantly,  in a realistic setup, energy can be evacuated from the probe not only  via electrons, but  it can also be transferred to the lattice phonons, which are in thermal equilibrium at a constant temperature~$T_0$. For the electron-phonon coupling strength~$\Sigma\vol$, with the probe volume~$\vol$, the time-averaged energy current flowing from the electron to the phonon subsystem is given by 
\mbox{$
	\Sigma\vol
	\big[\avTp^{\raisebox{-2.5pt}{$\scriptstyle5$}}-T_0^5\big]
	$,}  
where~$\avTp$ is the time-averaged probe temperature~\cite{Wellstood1994Mar,Appleyard1998Oct,Giazotto2006Mar,Gasparinetti2011May,Jezouin2013Oct,Laitinen2014May}. 
In the following, we neglect fluctuations of the energy current to the phonon bath. 
As a result, the total thermal conductance~$\kappa$ of the probe involves both the electronic and the phonon contribution,~\mbox{$\kappa(T)=\kappa_\text{e}(T)+\kappa_\text{ph}(T)$}, 
\begin{equation}\label{eq:kappa_def}
\kappa_\text{e}(T)=D gLT
\quad
\text{and}
\quad
\kappa_\text{ph}(T)=5\Sigma\mathcal{V}T^4
, 
\end{equation}
with the Lorenz number \mbox{$L=(\pi\kB/e)^2/3$}. 
Depending on the material, low-dimensional systems serving as probes may be characterized by different values of the temperature exponents in the electron-phonon coupling (e.g., in the range from 3 to 5 for graphene~\cite{Laitinen2014May}).

In addition, also the readout of the probe temperature and electrochemical potential would in general  lead to energy losses. Nonetheless, for the sake of simplicity, we here assume this readout to be \emph{non-invasive}. The unavoidable energy-losses due to the readout can, however, be included into our description on the same footing as the energy outflow to the phonon bath.

Finally, the dynamic properties of the probe are ruled by its charge and heat capacitances. Specifically, the charge capacitance~$C$ corresponds to the total (electrochemical) capacitance,~\mbox{$C=(1/\Cg+1/\Cq)^{-1}$}, containing both a geometrical,~$\Cg$, and a quantum component, \mbox{$\Cq=e^2\nu$}, with $\nu$ standing for the probe density of states at the Fermi level. 
Moreover, the electronic heat capacity is temperature-dependent, \mbox{$\CE(T)=\nu (\pi\kB)^2 T/3$}.

\subsection{Relevant timescales}\label{sec:timescales}

Relations between the different characteristic timescales occurring in the problem are crucial for the understanding of the physics underlying the proposed detection scheme, and set the validity of the theoretical approach to be applied. Below, we introduce all relevant timescales, and discuss relations between these timescales which need to be satisfied, so that the readout of the temperature and electrochemical potential fluctuations in the probe are feasible.

To begin with, the first important timescale to be considered is the electron-electron interaction time~$\tauee$, associated with internal thermalization of the probe electron gas. As mentioned above, a fast thermalization guarantees a Fermi distribution of the probe electrons, with a well defined temperature~$\Tp(t)$ at all times~$t$. Since this is a crucial condition for the readout of the probe temperature, we here assume $\tau_\text{e-e}$ to be much shorter or at most  of the order of the timescale at which electrons are emitted from the source~\cite{Gasparinetti2011May}. Electron-electron relaxation times were found to be of the order of 1-10 ns in metals~\cite{Pierre2003Aug} and 1ns in semiconductors~\cite{Katine1998Jan,Ferrier2004Dec}.

Next, we recall that the fluctuations in the electrical charge and energy current from the source occur on the timescale~$\perT$. When the fluctuating charge and energy currents flow into the probe, they give rise to fluctuations  of the electrochemical potential~$\mup(t)$ and of the temperature~$\Tp(t)$ in the probe. These electrochemical-potential and temperature fluctuations occur on timescales set by the probe charge and energy relaxation times, $\tauRC=C/(gD)$ and \mbox{$\tauE=\CE/\kappa$}, respectively. While single-electron sources are typically operated in the GHz regime~\cite{Feve2007May,Dubois2013Oct}, (metallic)  probes are expected to exhibit much slower dynamics in the MHz frequency range~\cite{Jezouin2013Oct}.  Improvements of this time can be reached by increasing the probe capacitance via the area and/or the volume of the probe or by tuning the transmsission of the QPC.

This motivates the choice of our theoretical approach -- the Boltzmann-Langevin approach as described in Sec.~\ref{sec:formalism}. It is applicable as long as the characteristic time of the current source, $\perT$, is much smaller than the charge and energy relaxation times, implying that $\mup(t)$ and $\Tp(t)$ change only by a small amount per drive cycle. Taken together, the required hierarchy of timescales can be established
\begin{equation}
\tauee \lesssim \perT \ll \tauRC, \tauE
.
\end{equation}

Finally, in order to have detectable temperature fluctuations, it is important that the energy relaxation is not completely dominated by the coupling to the lattice phonons. To meet this condition, $\kappa_\text{e}$ must be comparable to or larger than $\kappa_\text{ph}$. In terms of the relaxation time due to the electron-phonon scattering, \mbox{$\taueph=\CE/\kappa_\text{ph}$}, the optimal regime for detecting the temperature fluctuations is reached for~\mbox{$\taueph \gg \tauE$}. This condition basically demands that the energy relaxation is sufficiently slow to enable detection of  energy fluctuations. Indeed, at sub-Kelvin temperatures, thermal relaxation times due to electron-phonon coupling in metals reach tens of microseconds~\cite{Gasparinetti2015Jan}.

\section{Method and formalism}\label{sec:formalism}

Our aim is to acquire knowledge about the average and fluctuations of the charge,~$\Is(t)$, and energy,~$\Is^E(t)$, currents generated by the source.
This is achieved by investigating the average and the fluctuations of the electrochemical potential~$\mup(t)$ and the temperature~$\Tp(t)$ in the probe, which are, in turn, connected to the average charge and energy currents, $\Ip(t)$ and $\Ip^E(t)$, as well as to their fluctuations \textit{in the probe}.
Furthermore, it turns out that it is actually the heat current in the probe, \mbox{$\Jp(t)= \Ip^{E}(t)+\mup(t) \Ip(t)/e$}, rather than the energy current, $\Ip^{E}(t)$, that is naturally associated with the fluctuations of the temperature in the probe. 

In general, the procedure allowing us to relate the quantities of interest mentioned above consists of three steps:
First, we relate the electrochemical potential~$\mup(t)$ and the temperature~$\Tp(t)$ to the charge and the energy stored in the probe and formulate the dynamical equations for the two latter quantities  by resorting to general conservation laws.
Next, we determine the averages $\avmup$ and $\avTp$ in terms of the average source currents,~$\avIs$ and~$\avIs^E$. 
Finally,  using the Boltzmann-Langevin approach, we establish a relation between  the fluctuations of these currents  and the much slower fluctuations of the \textit{macroscopic probe} variables.
The \textit{bare} source currents and their fluctuations, which in this section enter as independent variables, will later be evaluated for a specific system by employing a Floquet scattering theory, see Sec.~\ref{sec:example}.

\subsection{General probe conservation equations}

As a first step, we express the time-dependent charge~$Q(t)$ and energy~$E(t)$ in the probe in terms of $\mup(t)$ and $\Tp(t)$. Considering the probe potential with respect to the electrochemical potential of the left reservoir~($\muL\equiv0$), the relations take the form
\begin{subequations}\label{eq:QE}
	\begin{gather}\label{eq:Q}
	Q(t)
	=  
	-\frac{C}{e}\,\mu_\text{p}(t) 
	,  
	\\\label{eq:E}
	E(t)
	=
	\frac{C}{2e^2}\,\mup^2(t)
	+ 
	\frac{\CE(\Tp)\,\Tp(t)}{2}
	.
	\end{gather}	
\end{subequations}
Importantly, the charge and energy in the probe can change when the charge and energy currents, $\Ip(t)$ and $\Ip^E(t)$, flow into the probe from the current source. In this case, the overall charge and energy conservation imposes the following continuity equations
\begin{subequations}\label{eq:dQdE}
	\begin{gather}\label{eq:dQ}
	\frac{\intd Q(t)}{\intd t}
	=
	\Ip(t)
	,
	\\ \label{eq:dE}
	\frac{\intd E(t)}{\intd t}
	=
	\Ip^{E}(t)-\Sigma\vol \big[\Tp^5(t)-T^5_0\big]
	.
	\end{gather}
\end{subequations}
Here, the term $\Sigma\vol\big[\Tp^5(t)-T^5_0\big]$ captures the fact that an additional time-dependent energy outflow occurs also due to coupling of probe electrons to the lattice phonons. 

\subsection{Time averaged quantities}\label{sec:timeaverage}

In the next step, we evaluate the time-averaged electrochemical potential~$\avmup$ and temperature~$\avTp$ in the probe. Note that the time with respect to which these quantities are averaged must be much longer than the response time of the probe.
Since there is no long-time charge build-up in the probe, we conclude from the charge conservation relation~(\ref{eq:dQ}) that~\mbox{$\avIp=0$}. In consequence, the electrochemical potential in the probe must adjust so that this constraint is fullfilled. 
This is the detection principle of the probe: the built-up of the potential allows us to make conclusions about the charge current into the probe (and analogously from the temperature for the conservation of energy).

The total, time-averaged current flowing into the probe has the form
\begin{align}\label{eq:ipeq}
\avIp  
= \ & 
D\bigg[
\avIs 
+
\frac{e}{h}
\int\!\!\intd E\, 
\big\{f_\text{p}(E)-f_\text{L}(E)\big\}
\bigg] 
\nonumber\\
=\ & 
D\bigg[\avIs+g\frac{\avmup}{e}\bigg]
,
\end{align}
with 
\mbox{$
	f_\alpha(E)
	=
	\big\{1+\text{exp}\big[(E-\mu_\alpha)/(\kB T_\alpha)\big]\big\}^{-1}
	$}
denoting the Fermi function for the left reservoir~(\mbox{$\alpha=\text{L}$}) and the probe~(\mbox{$\alpha=\text{p}$}).
Moreover, the total current~(\ref{eq:ipeq}) consists of a direct contribution from the source, $\avIs$, and a current flowing out of the probe, due to the probe potential~$\avmup$. 
The prefactor~$D$ in Eq.~(\ref{eq:ipeq}) stems from the conservation of the particle current at the QPC.  Together with the condition of zero average charge current in the probe~(\mbox{$\avIp=0$}) discussed above, this yields
\begin{equation}\label{eq:muav}
\avmup
=
-\frac{e}{g}\avIs
.
\end{equation}
As a result, one can see that the time-averaged probe electrochemical potential~$\avmup$ is solely determined by the time-averaged source current~$\avIs$, and is independent of the QPC-transparency.
In general, the explicit form of this current relies on a specific implementation of the current source. In this paper, it is illustrated for the example of a time-dependently driven mesoscopic capacitor in Sec.~\ref{sec:example}, derived by means of the Floquet scattering matrix approach (see Appendix~\ref{app:transport_theory}). Importantly, in the present section, the current~$\avIs$ enters the formalism as a general variable.

Similar to the charge, there is also no long-time accumulation of energy in the probe. From the time average of the energy-current conservation law, Eq.~(\ref{eq:dE}), we thus obtain 
\begin{equation}\label{eq:avIE_aux}
\avIp^E
- 
\Sigma\vol 
\big[\avTp^{\raisebox{-2.5pt}{$\scriptstyle5$}}-T^5_0\big]
=
0
.
\end{equation}
Furthermore, the energy current conservation at the QPC implies that 
\begin{align}\label{eq:jpeq}
\avIp^E
=\ &
D\bigg[
\avIs^E 
-
\frac{1}{h}
\int\!\!\intd E E
\big\{f_\text{p}(E)-f_\text{L}(E)\big\}
\bigg]
\nonumber \\
=\ & 
D\bigg[
\avIs^E
- 
\frac{\avIs^2}{2g}
-
\frac{L g}{2} \big(\avTp^{\raisebox{-2.5pt}{$\scriptstyle2$}}-\TL^2\big) 
\bigg]
.
\end{align}
Three different contributions can be identified in the equation above: the bare contribution from the current source~($\avIs^E$), a contribution stemming from Joule heating~(\mbox{$\propto\avIs^2$}), and the last term representing the energy current induced by the temperature difference between the left reservoir and the probe. 
Again, the energy current~$\avIs^E$ associated with the source enters here  as a general input variable, and it is calculated for the example of a time-dependently driven mesoscopic capacitor in Sec.~\ref{sec:example}.
With  Eqs.~(\ref{eq:avIE_aux})-(\ref{eq:jpeq}), we write the equation from which the average probe temperature~$\avTp$ can be found,
\begin{equation}\label{eq:tempeq}
\avTp^{\raisebox{-2.5pt}{$\scriptstyle5$}}
+ 
\frac{D}{\Sigma\vol}\frac{Lg}{2}\,
\avTp^{\raisebox{-2.5pt}{$\scriptstyle2$}}
=
T^5_0
+
\frac{D}{\Sigma\vol} 
\bigg[
\avIs^E
-
\frac{\avIs^2}{2g}
+ 
\frac{L g}{2}\,
\TL^2
\bigg]
.
\end{equation}
Since this expression shows that~$\avTp$ is determined by the average source charge and energy currents, $\avIs$ and $\avIs^E$, the readout of this temperature -- together with the readout of the probe electrochemical potential~\mbox{$\avmup$ --}  serves as a means for the detection of the average source energy current. 
Furthermore, Eq.~(\ref{eq:tempeq}) also depends on the temperatures of the left reservoir, $\TL$, and of the phonon bath,~$T_0$, as well as on the ratio $D/(\Sigma\vol)$.  
Importantly, the QPC transmission~$D$ is easily tunable in experiments via gate voltages, and also the electron-phonon coupling strength can in principle be tuned, e.g., by adjusting either the electron density~\cite{Gasparinetti2011May,Appleyard1998Oct,Fong2013Oct} or the background temperature~\cite{Fong2013Oct}.  
We note that Eq.~(\ref{eq:tempeq}), which for metallic probes is a fifth-order equation in~$\avTp$, typically has to be solved numerically for a given source. 

\subsection{Fluctuations, Boltzmann-Langevin approach}

Finally, in the last step, we consider the fluctuations \emph{in time} using the Boltzmann-Langevin approach~\cite{Blanter2000Sep}. 
For this purpose, we begin by dividing the probe currents~$\Ip(t)$ and~$\Ip^E(t)$ into time-averaged parts and  fluctuating, time-dependent parts,
\begin{subequations}\label{eq:IJdec}
	\begin{gather}
	\Ip(t) = \avIp + \DIp(t)
	, 
	\\
	\Ip^E(t) = \avIp^E + \DIp^E(t)
	.
	\end{gather}
\end{subequations}
The time-averaged parts, $\avIp$ and $\avIp^E$, have been discussed in Sec.~\ref{sec:timeaverage}.
On the other hand, for the fluctuating parts, $\DIp(t)$ and $\DIp^E(t)$, we distinguish two distinct physical origins: (i) bare \emph{injector} fluctuations, $\dIp(t)$ and $\dIp^E(t)$, which arise from the source currents or from scattering at the QPC; (ii) the fluctuations in the probe currents induced by fluctuations of the probe electrochemical potential~$\mup(t)$ and of the probe temperature~$\Tp(t)$.
As a result, it becomes convenient to  decompose also the potential and temperature fluctuations into a constant and a fluctuating contribution,
\begin{subequations}\label{eq:muTdec}
	\begin{gather}
	\mup(t)= \avmup + \Dmup(t), 
	\\
	\Tp(t)= \avTp + \DTp(t).
	\end{gather}
\end{subequations}
We here make the crucial assumption that the fluctuations of~$\dIp(t)$ and~$\dIp^E(t)$ are much faster than the fluctuations of~$\Dmup(t)$ and~$\DTp(t)$, see Sec.~\ref{sec:timescales}, which enables a separation of timescales. In~other words, the probe is sufficiently large, so that the change in~$\mup(t)$ and~$\Tp(t)$ is small on the timescale of the particle injection, which allows us to expand all fluctuation-dependent quantities up to the linear order in~$\Dmup(t)$ and~$\DTp(t)$ -- note that the long-time changes in~$\mup(t)$ and $\Tp(t)$ do not necessarily need to be small for this expansion to be valid.
In consequence, we obtain
\begin{equation}\label{eq:flucteq}
\renewcommand{\arraystretch}{2}
\begin{pmatrix}
\DIp(t) \\ \DIp^E(t)
\end{pmatrix}
\!=\!
\begin{pmatrix}
\dIp(t) \\ \dIp^E(t)
\end{pmatrix}
\!+\!
\begin{pmatrix}
\frac{\partial\avIp}{\partial\avmup} & \frac{\partial\avIp}{\partial\avTp}
\\
\frac{\partial\avIp^E}{\partial\avmup} & \frac{\partial\avIp^E}{\partial\avTp}
\end{pmatrix}
\!\cdot\!
\begin{pmatrix}
\Dmup(t) \\ \DTp(t)
\end{pmatrix}
\!\!.
\end{equation}
The bare fluctuations, $\dIp(t)$ and $\dIp^E(t)$, constitute the \emph{Langevin sources} of fluctuations in our theoretical approach, and they are explicitly derived for the case of a time-dependently driven mesoscopic capacitor in Sec.~\ref{sec:example}.
The prefactors of the fluctuations~$\Dmup(t)$ and~$\DTp(t)$ in Eq.~(\ref{eq:flucteq}) can be straightforwardly calculated from Eqs.~(\ref{eq:ipeq}) and~(\ref{eq:jpeq}) for~$\avIp$ and~$\avIp^E$, respectively, 
\begin{equation}\label{eq:response_coeff}
\renewcommand{\arraystretch}{2}
\begin{pmatrix}
\frac{\partial\avIp}{\partial\avmup} & \frac{\partial\avIp}{\partial\avTp}
\\
\frac{\partial\avIp^E}{\partial\avmup} & \frac{\partial\avIp^E}{\partial\avTp}
\end{pmatrix}
=
Dg
\begin{pmatrix}
1/e & 0
\\
-\avmup/e^2 & -L\avTp
\end{pmatrix}
\!\!.
\end{equation}
Note that we are considering  a conductor which does not break electron-hole symmetry, which essentially means that the thermoelectric linear-response coefficients vanish, that is, \mbox{$\partial\avIp/\partial \avTp=0$} and  \mbox{$\partial\avJp/\partial\avmup=0$}.

Next, employing Eqs.~(\ref{eq:QE})-(\ref{eq:dQdE}), we relate the bare current fluctuations, $\dIp(t)$ and $\dIp^E(t)$, to the electrochemical potential~$\Dmup(t)$ and temperature~$\DTp(t)$ fluctuations of the probe. 
It turns out that the sought relations are especially convenient to derive in frequency space. By combining the Fourier transforms of Eqs.~(\ref{eq:QE})-(\ref{eq:dQdE}), and inserting these expressions into Eq.~(\ref{eq:IJdec}), we obtain formulas relating  the macroscopic probe variables $\Dmup(\omega)$ and $\DTp(\omega)$ to the bare fluctuations of the charge current~$\dIp(\omega)$ and the heat current,
$\dJp(\omega)=\dIp^E(\omega)+\avmup\dIp(\omega)/e$
which have the  form
\begin{gather}\label{eq:delta_I}
\frac{\Dmup(\omega)}{-e} 
=
\frac{1}{D g}\, 
\frac{\dIp(\omega)}{1+i\omega\tauRC}
,
\\\label{eq:delta_J}
\DTp(\omega)  
=
\frac{1}{\kappa}\, 
\frac{\dJp(\omega)}{1+i \omega\tauE } 
.
\end{gather}
Importantly, Eqs.~(\ref{eq:delta_I})-(\ref{eq:delta_J}) are the \textit{key input} for the realization of heat- and charge-current noise detection via potential and temperature fluctuations. 
Here, we defined $\kappa\equiv\kappa(\avTp)$ as the thermal conductance at the probe temperature~$\avTp$. Similarly, we define $C_E\equiv C_E(\avTp)$, which, together with $\kappa$, enters the energy relaxation time \mbox{$\tauE=\CE/\kappa$}. 

The analysis of Eqs.~(\ref{eq:delta_I}) and (\ref{eq:delta_J}) indicates that there is an effective time scale on which the potential and temperature fluctuations have to be small in order for the linearization performed in Eq.~(\ref{eq:flucteq}) to hold. We find these effective time scales to 
be
$\sqrt{1/\omega^2+\tauRC^2}$ [with the charge relaxation time \mbox{$\tauRC=C/(Dg)$}] and $\sqrt{1/\omega^2+\tauE^2}$.
In particular, this means that the fluctuations  in~$\Tp(t)$ and~$\mup(t)$ have to be small either on the timescales set by the probe response -- given by $\tauRC$ and~\mbox{$\tauE$ --} or on the timescale on which the fluctuations are probed, namely, $1/\omega$. 
From a practical point of view, this implies that even when the probe-response times diverge (for example, $\tauRC$ for \mbox{$D\rightarrow 0$}), linear relations between source current fluctuations and electrochemical potential and temperature fluctuations, Eqs.~(\ref{eq:delta_I})-(\ref{eq:delta_J}), can still be found at finite frequencies $\omega$.

\section{Electrochemical potential and temperature fluctuations}\label{sec:mup_Tp_fluct}

\subsection{Relation to probe current fluctuations}\label{sec:probefluct}

With the use of Eqs.~(\ref{eq:delta_I})-(\ref{eq:delta_J}), we can now express the charge-current noise, the heat-current noise and the mixed noise in terms of the (symmetrized) frequency-dependent correlators of electrochemical potential and temperature. The latter are given by
\begin{multline}\label{eq:corr_def}
\pi\delta(\omega+\omega')
\Pp^{XY}(\omega)
\\
=
\big\langle
\Delta X_\text{p}(\omega)\Delta Y_\text{p}(\omega')
+
\Delta X_\text{p}(\omega')\Delta Y_\text{p}(\omega)
\big\rangle_\text{stat}\ 
,
\end{multline}
with \mbox{$X,Y=T,\mu$}, and $\langle\dots\rangle_\text{stat}$ denoting the statistical average over  Langevin sources. Note that the delta function \mbox{$\delta(\omega+\omega')$} stems from the invariance of fluctuations with respect to translations in time. This is a valid assumption because the time-dependent driving of the source takes place on a much faster timescale than the fluctuations of the probe
quantities.\footnote{%
Recall that the large separation of timescales between the slow fluctuations of the probe properties and the fast, bare fluctuations of the currents has been used before when setting up the Boltzmann-Langevin equations, see Eq.~(\ref{eq:flucteq}).
}
The two auto-correlators $\Pp^{\mu\mu}(\omega)$ and  $\Pp^{TT}(\omega)$, as well as the mixed correlator~$\Pp^{\mu T}(\omega)$, all defined by Eq.~(\ref{eq:corr_def}), depend on the correlators of the bare fluctuations  $\delta I_\text{p}(\omega)$ and $\delta J_\text{p}(\omega)$, the Langevin sources, through Eqs.~(\ref{eq:delta_I})-(\ref{eq:delta_J}). 
They are equivalently written as 
\begin{multline}\label{eq:corr_def2}
\pi \delta(\omega+\omega')\Pp^{AB}(\omega)
\\
= 
\big\langle
\delta A_\text{p}(\omega)\delta B_\text{p}(\omega')
+
\delta A_\text{p}(\omega')\delta B_\text{p}(\omega)
\big\rangle_\text{stat}\ 
,
\end{multline}
for \mbox{$A,B=I,J$}. These bare source correlators, in the frequency-regime relevant for Eq.~(\ref{eq:corr_def}), are just given by the low-frequency, period-averaged limit,
\begin{equation}
\Pp^{AB}(\omega)
\approx
\Pp^{AB}(\omega=0)
\equiv
\Pp^{AB}
.
\end{equation}
This is because the macroscopic probe quantities, $\Tp(\omega)$ and $\mup(\omega)$ fluctuate on much smaller frequencies than the probe currents, $\delta A_\text{p}(\omega)$ and $\delta B_\text{p}(\omega)$, do.  
The probe currents~$\Ip(t)$ and~$\Jp(t)$ are typically of quantum mechanical nature. Thus, for the evaluation of the correlators~$\Pp^{AB}$ of Eq.~(\ref{eq:corr_def2}), these currents have to be replaced by the  corresponding operators~$\opIp(t)$ and~$\opJp(t)$, 
\begin{equation}\label{Pp_def}
\Pp^{AB} 
= 
\frac{1}{2}
\int\limits^{\perT}_{0}\!\frac{\intd t}{\perT} 
\int\limits^{\infty}_{-\infty}\!\!\intd t^\prime\,
\big\langle 
\big\{\delta \hat{A}_\text{p}(t), \delta \hat{B}_\text{p}(t+t^\prime)\big\} 
\big\rangle, 
\end{equation}
with~\mbox{$\left\{\ldots,\ldots\right\}$} standing for the anticommutator and~$\langle\ldots\rangle$ representing the quantum-statistical average. The explicit expressions for $\Pp^{AB}$ are calculated using the quantum-coherent Floquet scattering theory, see Sec.~\ref{sec:example} and Appendices~\ref{app:transport_theory}-\ref{app:meso_induced} for general expressions and further details of the derivation.

Combining Eqs.~(\ref{eq:corr_def})-(\ref{eq:corr_def2}), and then employing relations~(\ref{eq:delta_I})-(\ref{eq:delta_J}), we can write the frequency-dependent probe electrochemical potential and temperature correlators,~$\Pp^{\mu\mu}(\omega)$ and~$\Pp^{TT}(\omega)$, which are directly proportional to the corresponding charge and heat current correlators,~$\Pp^{II}$ and~$\Pp^{JJ}$,
\begin{gather}\label{eq:Pmumu}
\Pp^{\mu\mu}(\omega)
=
\frac{e^2}{(D g)^2} 
\dfrac{1}{1+ (\omega  \tauRC)^2}\,
\Pp^{II}
,
\\\label{eq:PTT}
\Pp^{TT}(\omega)
=
\frac{1}{\kappa^2}
\dfrac{1}{1+ (\omega\tauE)^2 }\, 
\Pp^{JJ}
.
\end{gather}
An analogous expression can be found also for the mixed correlator,
\begin{equation}\label{eq:PmuT}
\Pp^{\mu T}(\omega)
=
\frac{-e}{D g\kappa}
\frac{
	1+\omega^2 \tauRC \tauE
}{
	\big[1+ (\omega\tauE)^2\big]
	\big[1+ (\omega\tauRC)^2\big]
}\,
\Pp^{IJ}
.
\end{equation}
Here, the order of the superscript of this \emph{symmetrized} correlator is irrelevant, that is, \mbox{$\Pp^{\mu T}(\omega)\equiv\Pp^{T\mu}(\omega)$} and $\Pp^{IJ}\equiv\Pp^{JI}$. The three equations above establish the direct relation between charge- and \emph{heat}-current noise in the probe and the frequency-dependent fluctuations of its macroscopic quantities -- the temperature and electrochemical potential. These equations provide the basis for the proposed detection scheme, making the properties of charge and energy currents emitted from the source experimentally accessible. 

Nevertheless, in Eqs.~(\ref{eq:PTT})-(\ref{eq:PmuT}), information about the \textit{energy}-current fluctuations is still obscured. To extract the equivalent correlators for the (mixed) \emph{energy}-current fluctuations~($\Pp^{IE}$)~$\Pp^{EE}$, we invoke the relation between the heat and energy current fluctuations, $\dJp(\omega)=\dIp^E(\omega)+\avmup\dIp(\omega)/e$. [To keep the notation simple, we here write $E$ instead of $I^E$ in the superscripts of $\mathcal{P}$]. This allows us to find $\Pp^{IE}$ and~$\Pp^{EE}$ from the three correlators~$\Pp^{II}$, $\Pp^{JJ}$ and~$\Pp^{IJ}$ introduced above, by applying step by step the relations 
\begin{gather}
\Pp^{IJ}
= 
\Pp^{IE}
+
\frac{\avmup}{e}
\Pp^{II}
,
\\
\Pp^{JJ}
=
\Pp^{EE}
+
\frac{\avmup}{e}
\Pp^{IE}
+
\frac{\avmup^2}{e^2}
\Pp^{II}
,
\end{gather}
and employing the knowledge of the probe potential obtained from independent measurements.	

Furthermore, the frequency-dependence of the (mixed) potential and temperature fluctuations, given in Eqs.~(\ref{eq:Pmumu})-(\ref{eq:PmuT}), sets a fundamental constraint on the bandwidth of the measurement, that is, the smaller the relaxation times $\tauRC$ and $\tauE$, the less the bandwidth is limited. 
From the experimental point of view, an increased bandwidth is highly desirable as it allows for a measurement speed-up, which is particularly important when a large number of measurements has to be performed in order to guarantee an optimal subtraction of a fluctuating background signal.

Indeed, the subtraction of a background signal turns out to be crucial when one aims at extracting the desired fluctuations of the \textit{source} currents from the \textit{probe}-current fluctuations. 
How this can be done is shown in the following correlator decomposition, which for simplicity is performed on the correlators $\mathcal{P}^{AB}_\text{p}$, for $A,B=I,J$, since the frequency dependence of  Eqs.~(\ref{eq:Pmumu})-(\ref{eq:PmuT}) is not crucial for this decomposition.

\subsection{Correlator decomposition}\label{sec:decompose}

In the previous section, we have demonstrated how fluctuations of the temperature~$\Tp(t)$ and electrochemical potential~$\mup(t)$ in the probe are related to the relevant charge-current~($\Pp^{II}$) and heat-current~($\Pp^{JJ}$) auto-correlators, as well as to the mixed-current~($\Pp^{IJ}$) correlator \emph{in the probe}.  
However, in order to acquire a complete understanding on how the sought \emph{source fluctuations} are influenced by the detection -- or in other words, how the correlators of the probe currents are linked to the correlators of the source currents -- it is instructive to divide~$\Pp^{AB}$ into terms with physically distinct origins.  

The first step involves writing~$\Pp^{AB}$ as a sum of two terms: the first~($\Po^{AB}$) arising already in the absence of a working source, and the second~($\Ps^{AB}$) that can be attributed to the source driving,
\begin{equation}\label{eq:Ptotal}
\Pp^{AB}
=
\Po^{AB}
+
\Ps^{AB}
.
\end{equation}
While $\Po^{AB}$ is due to the non-zero, constant temperatures of the left reservoir and the probe, $\Ps^{AB}$ contributes only if the source is on and is equal to zero otherwise.
Consequently, when evaluating~$\Po^{AB}$, the probe effectively acts as a second, right~(R), reservoir characterized by a distribution function 
\mbox{$
	f_\text{R}(E)
	=
	\big\{1+\text{exp}\big[E/(\kB\TR)\big]\big\}^{-1}
	$,}
with the temperature~$\TR$ determined from Eq.~(\ref{eq:tempeq}). 
In an experiment, this contribution is obtained from a noise measurement carried out when the time-dependent driving is switched off. In order to obtain $\Ps^{AB}$, one has to subtract~$\Po^{AB}$ from results of subsequent measurements of $\Pp^{AB}$ with a working source.

In the second step, the fluctuations attributed to the driving source~$\Ps^{AB}$ can be further separated into direct~($\Psdir^{AB}$) and induced~($\Psind^{AB}$) correlations as
\begin{equation}\label{eq:Ps}
\Ps^{AB}
=
\Psdir^{AB}
+
\Psind^{AB}
.
\end{equation}
The direct correlations are the ones obtained, if the probe had been kept at constant temperature and electrochemical potential, with the distribution function~$f_\text{R}(E)$. It essentially means that these correlations arise due to the source driving in the absence of voltage build-up and temperature raise induced by the detection itself. In fact, it is precisely this term that the detection scheme proposed in this paper aims to extract. In order to obtain~$\Psdir^{AB}$ theoretically from $\Ps^{AB}$, the temperature and electrochemical potential have to be set to their equilibrium values~$\avTp$ and~$\avmup$, see Eq.~(\ref{corrFloq}). 
On the other hand, the remaining term~$\Psind^{AB}$ corresponds to the induced correlations, which originate from the fluctuations of the probe properties,~$\mup(t)$ and~$\Tp(t)$, induced by the source driving. Importantly, these correlations represent the back-action effect resulting from the measurement itself.\footnote{Note that in \textit{more complex} multi-terminal setups backaction can be partly suppressed if the system is at the same time also chiral~\cite{Parmentier2012Apr}.}
In an experiment, such a separation into induced and direct parts of the source fluctuations is not possible. For this reason, it is important to analyze the relative magnitude of these terms as we do for the example of a mesoscopic capacitor acting as the current source in the following section.

\section{Time-dependently driven mesoscopic capacitor}\label{sec:example}

\begin{figure}[tb]
	\centering
	\includegraphics[width=3in]{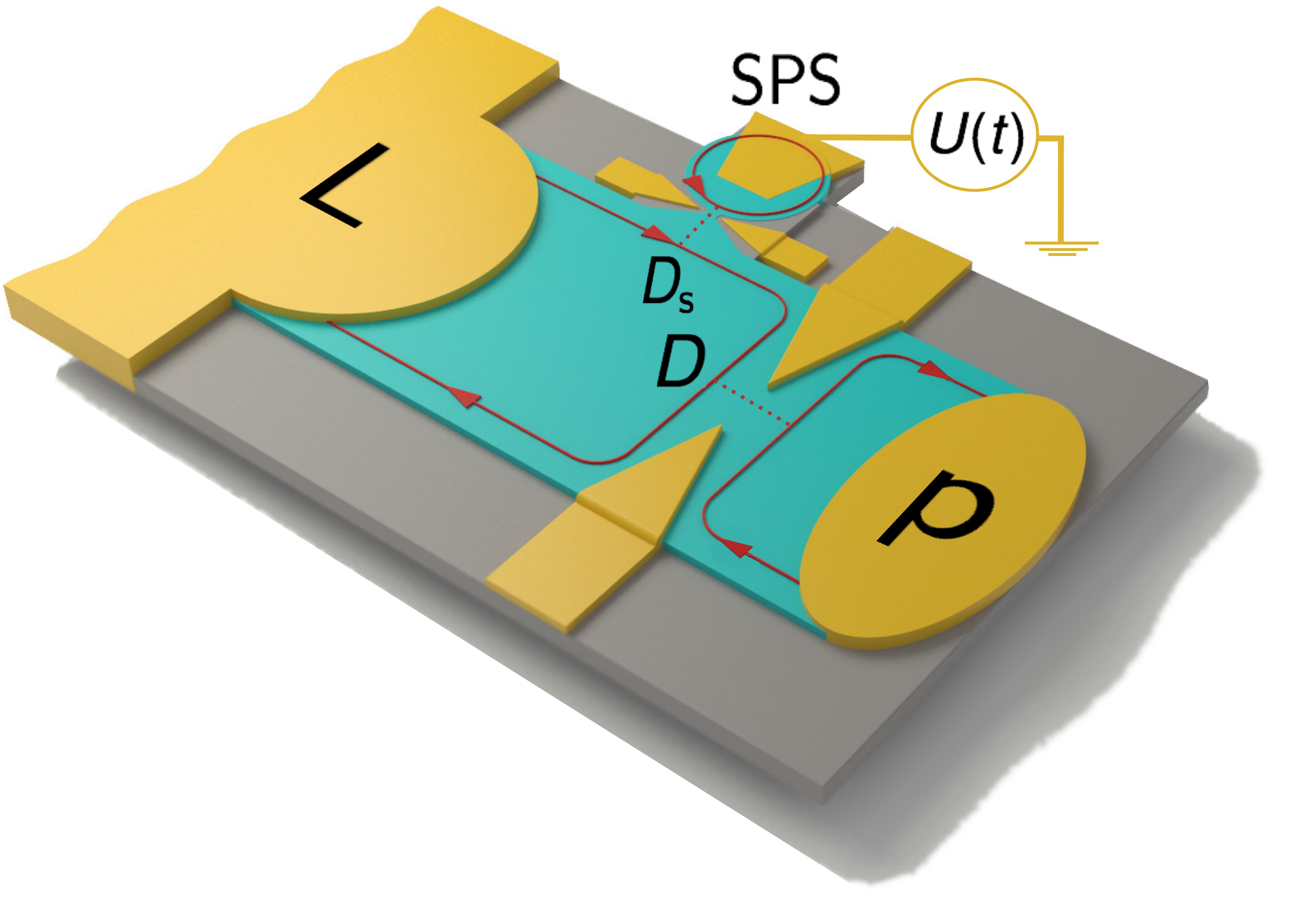}
	\caption{
		Edge-state implementation of the injector-detector system with the single-particle source~(SPS) that consists of a mesoscopic capacitor slowly driven by a time-periodic potential~$U(t)$. 
		The capacitor is connected to the edge state via a QPC with transparency~$\Ds$. The transport along the edge states is chiral, with the direction of transport shown by arrows. 
	}
	\label{fig:example-sys}
\end{figure}

We now provide a proof-of-principle of our approach and consider a specific time-dependent injector source generating a quantized, alternating current. 
We assume that the total injector-detector system is implemented with edge states in a two-terminal conductor in the integer quantum Hall regime with filling factor one, as shown in Fig.~\ref{fig:example-sys}. 
In particular, the injector part of our exemplary system consists of a mesoscopic capacitor, as realized experimentally by F\`{eve}~\etal~\cite{Feve2007May}. Due to the confinement of charge in the mesoscopic capacitor plate, the source has a discrete energy spectrum characterized by the level spacing~$\Delta$. This source is side-coupled to the active edge state via a QPC with transparency~$\Ds$. By applying a time-periodic potential~$U(t)$ to the capacitor, under ideal conditions (specified below), the capacitor emits exactly one electron and one hole per driving period~$\perT$.
Moreover, both the left reservoir~(\mbox{$\alpha=\text{L}$}) and the probe~(\mbox{$\alpha=\text{p}$}) are described by Fermi functions~$f_\alpha(E)$, with electrochemical potentials~$\mu_\alpha$ and temperatures~$T_\alpha$. 
Recall that the left reservoir -- namely, the terminal at the injector side -- is assumed throughout the paper to be grounded~(\mbox{$\muL=0$}) and its temperature equals the phonon temperature~(\mbox{$\TL=T_0$}), and so does the probe in the case when the source driving is switched off. This means, in turn, that there is no direct thermal bias, except the one induced by the detection. 

We describe the coherent conductor shown in Fig.~\ref{fig:example-sys} by the time-dependent (Floquet) scattering matrix approach. Since the system is periodically driven in time, the scattering matrix elements, $S_{\alpha\beta}(E,t)$ for~\mbox{$\alpha,\beta=\text{L},\text{p}$}, depend in general both on the initial time and the final energy of the scattering process (or equivalently on two times or on two energies~\cite{Moskalets2011Sep}).
For the scatterer considered in Fig.~\ref{fig:example-sys}, with a forward-scattering mesoscopic capacitor coupled to one edge channel, connected in series with a QPC with energy-independent transparency~$D$, one can write the components of the time- and energy-dependent scattering matrix as
\begin{equation}
\begin{aligned}\label{eq:mescondamp}
S_\text{LL}(E,t)&=\sqrt{1-D}\ S(E,t),
\\ 
S_\text{pL}(E,t)&=\sqrt{D}\ S(E,t), 
\\
S_\text{Lp}(E,t)&=\sqrt{D},
\\  
S_\text{pp}(E,t)&=-\sqrt{1-D}. 
\end{aligned}
\end{equation}
Here,~$S(E,t)$ is the scattering matrix of the mesoscopic capacitor only.

\subsection{Adiabatic-response to slow driving}

Let us consider the adiabatic-response to a slow driving potential~$U(t)$ of the mesoscopic capacitor. This implies that the period of the drive~\mbox{$\perT=2\pi/\Omega$} is much larger than the temporal width~$\sigma$ of the wave packets associated with the particles emitted by the source. The width~$\sigma$ is a function of the transparency~$\Ds$, the level spacing~$\Delta$ and the driving potential~$U(t)$~\cite{Moskalets2011Sep}, and is treated here as an input parameter.
In addition, we consider the regime where the thermal energies of injector and probe terminals, $\kB T_0$ and $\kB\avTp$, are much smaller than the energy scale $\hbar/\sigma$ over which the scattering properties vary 
appreciably.\footnote{%
We note that the driving frequency~$\Omega$ is typically of the order of GHz~($\hbar\Omega\sim\mu\text{eV}$)~\cite{Feve2007May}, while temperatures~$T_0$ and~$\avTp$ are in the range of tens of~mK ($\kB T_0,\kB\avTp\sim\mu\text{eV}$)~\cite{Jezouin2013Oct}. This basically means that we require~$\hbar/\sigma$ to be of the order of hundreds of~$\mu$eV, which corresponds to a pulse width~$\sigma$ of tens of~ps. }
Consequently, we define the driving regime of interest by the following condition
\begin{equation}\label{eq:ad_cond}
\hbar \Omega,\ 
\kB T_0,\
\kB\avTp
\ll 
\frac{\hbar}{\sigma}
.
\end{equation}
In this adiabatic-response limit the scattering properties of the mesoscopic capacitor are characterized by the \textit{frozen} scattering matrix~$S_0(E,t)$, where the time~$t$ only enters parametrically. The low temperature condition further allows us to evaluate the scattering matrix at the Fermi energy of the left terminal, $E=0$. This gives for~\mbox{$S(t)\equiv S_0(0,t)$}, up to irrelevant phase factors~\cite{Moskalets2011Sep},
\begin{equation}
S(t)
=
\left\{
\begin{aligned}
&\dfrac{t-t_\text{e}+i\sigma}{t-t_\text{e}-i\sigma}
\quad
\text{for}
\quad 
0 \leq t < \perT/2 ,
\\[3pt]
&\dfrac{t-t_\text{h}-i\sigma}{t-t_\text{h}+i\sigma}
\quad
\text{for}
\quad
\perT/2 \leq t < \perT
,
\end{aligned}
\right.
\end{equation}
where the driving is such that only a single energy level participates in the emission process. Here, the times for electron~($t_\text{e}$) and hole~($t_\text{h}$) emission, defined by the capacitor and drive potential properties~\cite{Moskalets2011Sep}, are assumed to be well inside the respective halves of the period, so that temporal overlap of the emitted particle wave functions can be neglected.

In the energy description, the scattering matrix takes the general form
\begin{equation}
S(E_n,E_m)
=
\int\!\frac{\intd t}{\perT}\,
\text{e}^{i(n-m)\Omega t}
S(E_n,t)
,
\end{equation}
with~\mbox{$E_n=E+n\hbar\Omega$} and $S(E_n,E_m)$ denoting the Floquet scattering matrix. Here, \mbox{$n-m\in\mathbb{Z}$} is the number of Floquet energy quanta picked up during a scattering process.
In the energy description, the adiabatic condition corresponds to the assumption that the scattering properties are energy independent on the scale of~$N_\text{max}\hbar \Omega$, where $N_\text{max}$ stands for the maximum number of quanta absorbed or emitted in the scattering process at the capacitor. The corresponding Floquet scattering matrix amplitudes then only depend on the difference in energies, \mbox{$S(E_n,E_m)=S(E_n-E_m) \equiv S_{n-m}$}, and up to the leading order in $\sigma\Omega$, one 
finds\footnote{%
	The Floquet matrix is unitary, implying the relation \mbox{$\sum_{n=-\infty}^{\infty} S^*(E_n,E_m)S(E_n,E_q)=\delta_{mq}$}, which here is required to hold up to the first order in $\sigma\Omega$, only.} 
\begin{equation}\label{eq:scattering}
S_{n}
=  
\begin{cases} 
-2 \Omega \sigma e^{-n\Omega \sigma}  e^{in\Omega t_{\text e}} 
&
\text{for}\quad  n>0 
,
\\ 
-2 \Omega \sigma e^{n\Omega \sigma} e^{in\Omega t_{\text h}}
& 
\text{for}\quad  n<0
,
\\
\delta_{n,0} 
&  
\text{for}\quad  n=0
.
\end{cases}
\end{equation}
With this expression at hand, we can evaluate all quantities of interest, that is, the average values of charge and heat currents, as well as the relevant current correlators. See  Appendix~\ref{app:transport_theory} for the outline of the derivation and specific formulas.

\subsection{Average potential and temperature}

In order to determine the average temperature~$\avTp$ and electrochemical potential~$\avmup$ arising in the probe due to the driving, the average charge and energy currents have to be explicitly calculated using Eqs.~(\ref{Iav})-(\ref{IEav}) from Appendix~\ref{app:transport_theory} for the scattering matrix given in Eq.~(\ref{eq:mescondamp}) and Eq.~(\ref{eq:scattering}).
The average source charge current is \mbox{$\avIs=0$}, which comes as a consequence of the fact that the same number of electrons and holes is emitted by the source per period. Therefore also  the average electrochemical potential of the probe, given by Eq.~(\ref{eq:muav}), is zero, 
\begin{equation}
\avmup=0.
\end{equation}
Furthermore, the average energy current emitted by the source is \mbox{$\avIs^{E}=2\langle \epsilon \rangle/\perT$}, where the average energy of each emitted particles is \mbox{$\langle \epsilon \rangle=\hbar/(2\sigma)$} and the factor 2 stems from the fact that both an electron and a hole are emitted per period. 
Inserting this expression together with~\mbox{$\avIs=0$} into Eq.~(\ref{eq:tempeq}), we obtain an equation allowing us to find the average probe temperature,
\begin{equation}\label{eq:ex-T}
\bigg(\frac{\avTp}{T_0}\bigg)^{\!\!5}-1
=
\frac{5}{2}
\frac{\kappae}{\kappaph}
\Bigg[
1
-
\bigg(\frac{\bar T_\text p}{T_0}\bigg)^{\!\!2}
+
\frac{4\langle \epsilon \rangle}{\mathcal{T}gLT_0^2}
\Bigg]
.
\end{equation}

\begin{figure}[t!!!]
	\includegraphics[width=3in]{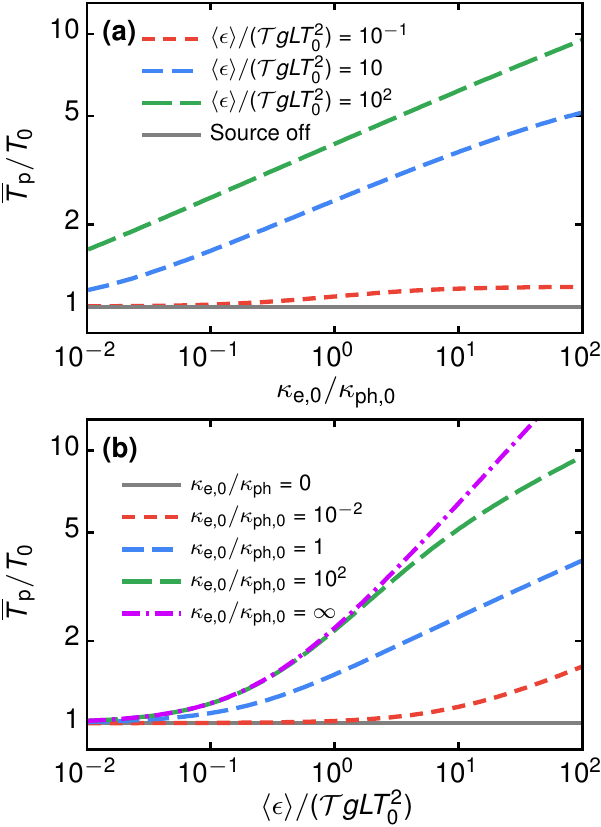}
	\caption{
		(a) Renormalized probe temperature $\avTp/T_0$ shown on a logarithmic scale as a function of the ratio of the electron and phonon thermal conductances $\kappae/\kappaph$ for selected values of the average energy of an emitted particle per period,~$\langle \epsilon \rangle/\perT$, scaled to the electronic heat current response of the probe,~$gL T_0^2$. 
		(b) Dependence of the renormalized probe temperature $\avTp/T_0$ (plotted logarithmically) on the ratio $\langle \epsilon \rangle/(\perT gL T_0^2 )$ for different values of $\kappae/\kappaph$. 
	}  
	\label{fig:T}
\end{figure} 

Here $\kappae\equiv\kappa_\text{e}(T_0)$ and $\kappaph\equiv\kappa_\text{ph}(T_0)$ are the electronic and the phonon thermal conductivity at the base temperature. The expression $gLT_0^2$ is the pure electronic heat response of the probe at the base temperature; it is equal to $\kappae T_0$ when $D=1$. The numerical solution for Eq.~(\ref{eq:ex-T}) is presented in Fig.~\ref{fig:T}. 
Specifically, panel~(a) shows the (renormalized) average probe temperature $\avTp/T_0$ as a function of the ratio between thermal conductances~$\kappae$ and~$\kappaph$. When the source is switched off (solid line), which effectively corresponds to~\mbox{$\langle \epsilon \rangle=0$}, the probe temperature is naturally always given by the phonon temperature~$T_0$. 
On the other hand, the larger the energy emitted from the source, the stronger is the deviation of the probe temperature from the phonon temperature. Also the  monotonic increase of~$\avTp/T_0$ shown in Fig.~\ref{fig:T}(a) is intuitively clear, since the increase of the temperature is expected to be larger in the limit when the energy loss to the phonon bath is small, that is, for~\mbox{$\kappae/\kappaph\gg1$}.

The dependence of the probe temperature on the energy emitted per particle in a period, $\langle \epsilon \rangle/\perT$ (normalized with respect to the electronic probe heat current $gLT_0^2$), is shown in panel (b) of Fig.~\ref{fig:T}. As mentioned above, the monotonic increase of the temperature fits in with the expectation that the value of the probe temperature is a signature of the  energy transported into the probe by the particles emitted from the source. This is, however, true only as long as the coupling to the phonon bath is not too strong: in the opposite limit of very strong coupling~(\mbox{$\kappae/\kappaph\rightarrow0$}), the probe temperature is always equal to the phonon temperature independently of the source driving, see the solid line in Fig.~\ref{fig:T}(b). In contrast, if the coupling to the phonon bath is negligibly small (dotted-dashed line), the average temperature in the probe is given by
\begin{equation}
\frac{\avTp}{T_0}
=
\sqrt{
	1
	+
	\frac{4\expec{\epsilon}}{\mathcal{T}gLT_0^2}
}
\quad
(\text{for}\ \kappaph\ll\kappae)
.
\end{equation}

Finally, we note that while the general solution of Eq.~(\ref{eq:ex-T}) for temperatures $\avTp$ can only be obtained numerically, we can still estimate the linear response, corresponding to the limit of~\mbox{$\avTp-T_0\ll T_0$}. In this regime, the drive has a small effect on the probe temperature, such that the analytic formula for the renormalized deviation from the bath temperature is given by  
\begin{equation}
\frac{\avTp-T_0}{T_0}
=
\frac{2\langle \epsilon \rangle}{\mathcal{T}gLT_0^2}
\bigg(1+\frac{\kappaph}{\kappae}\bigg)^{\!\!-1}
.
\end{equation}

\subsection{Potential and temperature fluctuations}\label{sec:meso_fluct}

We now turn our attention to the quantities of main interest in this paper, that is, the fluctuations of the probe temperature and electrochemical potential, $\Pp^{XY}(\omega)$, with~\mbox{$X,Y=\mu,T$}. 
Their frequency-dependent relation to the correlators of the probe charge and heat currents has been established in Sec.~\ref{sec:probefluct}, see Eqs.~(\ref{eq:Pmumu})-(\ref{eq:PmuT}). 
In order to obtain explicit expressions for the example of a slowly-driven mesoscopic capacitor discussed in this section, we have to derive correlators~$\Pp^{AB}$ for~\mbox{$A,B=I,J$}. Starting from the general expression given in Appendix~\ref{app:transport_theory}, this means that the following formula needs to be evaluated
\begin{widetext}
	\begin{align}\label{app_gencorr}
	\hspace*{-2pt}
	\Pp^{AB}=\ &
	D(1-D)\frac{1}{h} 
	\int\!\! \intd E\,
	x_A  x_B 
	\sum_n|S_{n}|^2
	\Big\{
	\fL(E_n)\big[1-\fp(E)\big]+\fp(E)\big[1-\fL(E_n)\big]
	\Big\}
	\nonumber \\ 
	&+
	D^2\frac{1}{h} 
	\int\!\! \intd E\, 
	\Big\{ 
	x_A x_B \fp(E)\big[1-\fp(E)\big]   
	+
	\sum_{nkq}
	\frac{x_Ax_{B,n}+x_{A,n}x_B}{2}
	S_{-q}^*S_{n-q}^{}S_{-k}^{}S_{n-k}^*\,
	\fL(E_q)\big[1-\fL(E_k)\big] 
	\Big\}
	,
	\end{align}
\end{widetext}
where the notation~$x_{A,n}$ should be read as: \mbox{$x_{I,n}= x_{I}\equiv e$} and \mbox{$x_{J,n}\equiv E_n-\avmup$}  (with \mbox{$x_J\equiv x_{J,0}$}).
Straight\-forward calculations employing the scattering matrix given in Eq.~(\ref{eq:scattering}) yield the desired expressions for the correlator components~$\Po^{AB}$,  $\Psdir^{AB}$ and  $\Psind^{AB}$, introduced in Sec.~\ref{sec:decompose}, which will be discussed in the remaining part of this section. 
For explicit derivations of~$\Po^{AB}$ and~$\Psdir^{AB}$, we refer the reader, e.g.,  to Ref.~\cite{Battista2014Dec,Crepieux2014Dec} where the equilibrium correlations, corresponding to~$\Po^{AB}$, are calculated, while more details regarding the derivation of the direct contribution~$\Psdir^{AB}$ will be presented in Ref.~\cite{Dashti17}. 
On the contrary, the evaluation of the induced part~$\Psind^{AB}$, which is unique for the type of measurement we are proposing here, is described in more detail in Appendix~\ref{app:meso_induced}.
For the sake of compactness of the following discussion, we introduce the auxiliary function~$\funH_q(\avTp/T_0)$ defined as:
\begin{equation}\label{ex:Hq}
	\funH_q\bigg(\frac{\avTp}{T_0}\!\bigg)
	=
	\int\!\!  \frac{\intd E E^q}{(k_\text B T_0)^{q+1}} 
	\big[\fL(E)-\fp(E)\big] \fL(E). 
\end{equation}
%

\subsubsection{Electrochemical potential correlations vs. charge-current noise}\label{sec:mumu_correl_ex}

We begin with the analysis of the electrochemical potential fluctuations~$\Pp^{\mu\mu}(\omega)$ as a measure of the charge-current correlations, which are well-known and have been measured previously~\cite{Mahe2010Nov,Parmentier2012Apr,Bocquillon2012May,Gabelli2013Feb,Dubois2013Oct}.
Still, we use this correlator to demonstrate here the working principle of our injector-detector scheme for the example of a time-dependently driven mesoscopic capacitor acting as the current source.  
Starting from Eq.~(\ref{eq:Pmumu}) with Eq.~(\ref{app_gencorr}), and making use of the correlator decomposition discussed above, Eqs.~(\ref{eq:Ptotal})-(\ref{eq:Ps}), we can write
\begin{equation}\label{eq:Pmumu_ex}
\frac{\Pp^{\mu\mu}(\omega)}{\Po^{II}}
=
\frac{e^2}{(gD)^2} \frac{1}{1+(\omega\tauRC)^2} 
\left[
	1+ \Fsdir^{II}+\Fsind^{II}
\right].
\end{equation}
Here, \mbox{$\Po^{II}=2Dgk_\text{B}T_0$} corresponds to the well-known equilibrium thermal noise of the QPC, which is related to the QPC conductance~$Dg$ through the fluctuation-dissipation theorem, see e.g.~\cite{Blanter2000Sep}. 
Moreover, the direct and induced components of the renormalized, dimensionless source charge-current correlators, that is, \mbox{$\Fsdir^{II}\equiv\Psdir^{II}/\Po^{II}$} and \mbox{$\Fsind^{II}\equiv\Psind^{II}/\Po^{II}$}, respectively,  are given by
\begin{gather}\label{eq:FII_dir_ex}
\Fsdir^{II}
=
(1-D)\frac{\hbar\Omega}{\kB T_0}
,
\\\label{eq:FII_ind_ex}
\Fsind^{II}
= 
\frac{D \big(\avTp- T_0\big)}{2T_0}
+
(1-D)\,\funH_0\bigg(\frac{\avTp}{T_0}\!\bigg)
.
\end{gather}
%

\begin{figure}[t!!!]
	\includegraphics[width=3in]{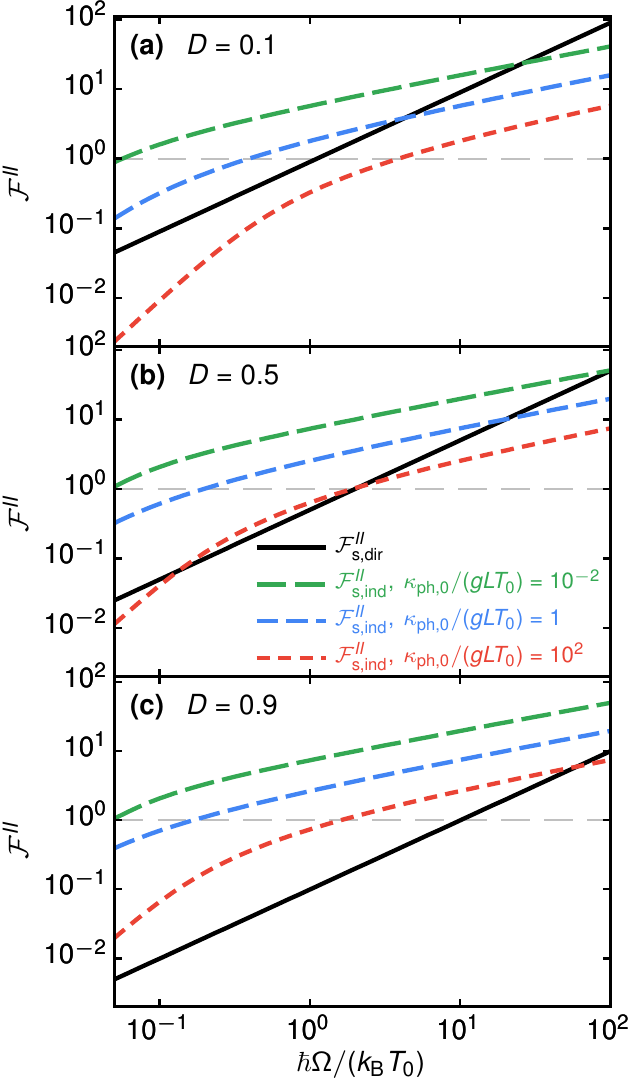}
	\caption{%
		Direct ($\Fsdir^{II}$, solid line) and induced ($\Fsind$, dashed lines) components of the renormalized, dimensionless source charge-current correlators shown as a function of $\hbar \Omega /(\kB T_0)$ for several values of the normalized electron-phonon coupling $\kappa_{\text{ph},0}/(gLT_0)$.
		The horizontal (finely dashed) line  represents the background noise, with respect to which the expressions are normalized.
		Note that only~$\Fsind^{II}$ depends on this coupling, as can be seen from Eqs.~(\ref{eq:FII_dir_ex})-(\ref{eq:FII_ind_ex}).
		Different panels correspond to specific values of the QPC transparency: (a)~\mbox{$D=0.1$}, (b)~\mbox{$D=0.5$} and (c)~\mbox{$D=0.9$}. 
		The pulse width in all panels is~\mbox{$2\Omega \sigma=0.001$}. 
		Finally, the lower-bound cutoff for values of~$\hbar \Omega /(\kB T_0)$ is imposed by the adiabatic condition, Eq.~(\ref{eq:ad_cond}), requiring \mbox{$\hbar \Omega /(\kB T_0)\gg\sigma \Omega $}.
	}
	\label{fig:FII}
\end{figure}

The aim of the present analysis is to investigate the ratio between the induced and direct contributions to the fluctuations in order to demonstrate the feasibility of the detection scheme proposed in this manuscript.
For this purpose, we fix the direct fluctuations~$\Fsdir^{II}$ to the ideal, simple regime, described by Eq.~(\ref{eq:FII_dir_ex}). In a realistic experiment, one might instead be interested in studying the direct noise as a function of the driving signal shape~\cite{Gabelli2013Feb,Dubois2013Oct,Misiorny17} or other source-related properties deviating from the ideal-driving case presented here.

The direct part~$\Fsdir^{II}$, which in this regime only depends on the QPC transmission and the number of particles impinging on the QPC per period~\cite{Olkhovskaya2008Oct}, is shown as the solid line in Fig.~\ref{fig:FII}. The linear dependence of~$\Fsdir^{II}$ on the transmission probability~$D$ is visible in Fig.~\ref{fig:FII}, where the magnitude of~$\Fsdir^{II}$ is increased by approximately one order when diminishing the transmission from~\mbox{$D=0.9$} (bottom panel) to~\mbox{$D=0.1$} (top panel). In contrast, this part of the noise is inherently insensitive to the probe properties, quantified by the ratio $\kappaph/\kappa_0$.

The situation is different for the induced part~$\Fsind^{II}$, which depends on the detector properties through the induced temperature~$\avTp$, see Eq.~(\ref{eq:FII_ind_ex}). 
Since, the transmission~$D$ now enters the noise also indirectly via the fraction $\kappa_{\text{e},0}/\kappa_{\text{ph},0}=DgLT_0/\kappa_{\text{ph},0}$, we choose in Fig.~\ref{fig:FII} to show different curves for  selected values of the \emph{probe} parameter~\mbox{$\kappaph/(gLT_0)$}, only, whereas different panels represent indicated values of~$D$.
It can be seen that the rise of the probe temperature~$\avTp$ with decreasing $\kappa_{\text{ph},0}$, demonstrated in Fig.~\ref{fig:T}(b), is also accompanied by an increase of $\Fsind^{II}$.
On the other hand, the dependence of the induced part on the transmission probability~$D$ is negligible, as compared to the $D$-dependence of the direct part. The reason for this is that the magnitudes of the two contributions, \mbox{$(\avTp-T_0)/(2T_0)$} and $-\funH_0(\avTp/T_0)$, to $\Fsind^{II}$, Eq.~(\ref{eq:FII_ind_ex}), are of similar order. Therefore, the terms proportional to~$D$ approximately cancel each other, such that \mbox{$\Fsind\approx-\funH_0(\avTp/T_0)$}. Note that~$\Fsind^{II}$ still weakly depends on~$D$ through~$\avTp$.

Consequently, in an experiment, where the transmission probability~$D$ can be easily tuned, the relative magnitude of~$\Fsdir^{II}$ and~$\Fsind^{II}$ can be adjusted as well. 
Indeed, the three panels of Fig.~\ref{fig:FII} clearly illustrate that, while at \mbox{$D=0.9$} [see Fig.~\ref{fig:FII}(c)] the desired direct signal is fully covered by the induced fluctuations, a small QPC transmission $D=0.1$ [see Fig.~\ref{fig:FII}(a)] allows for a detection of charge-current fluctuations of the source. At $D=0.1$ and $\kappa_{\text{ph},0}/(gLT_0)=10^{2}$, as shown in panel (a) of Fig.~\ref{fig:FII}, the sought-for, direct part of the charge-current fluctuations is always much larger than the induced noise, as required for a measurement of the former. A detailed discussion of the ideal choice of tunable parameters is presented in Sec.~\ref{sec:optimization}.

\subsubsection{Temperature correlations vs. heat-current noise}\label{sec:TT_correl_ex}

\begin{figure}[t!!!]
	\centering
	\includegraphics[width=3in]{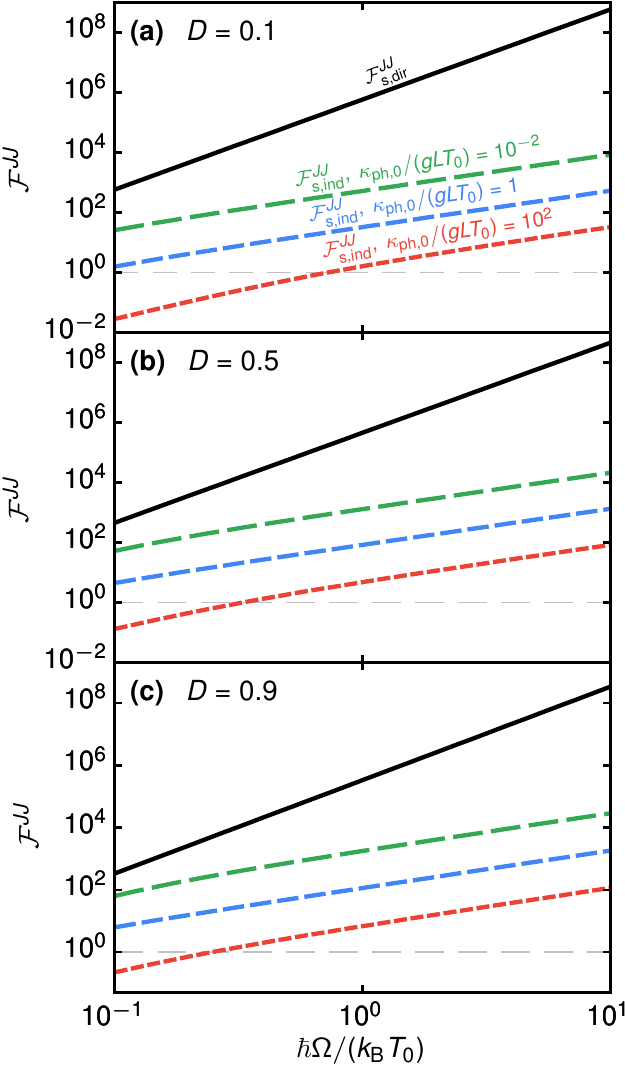}
	\caption{
	Analogous to Fig.~\ref{fig:FII} except that now direct (solid line) and induced (dashed lines) components of the normalized, dimensionless source heat-current correlators are shown as a function of $\hbar \Omega / (\kB T_0) $ for selected values of the electron-phonon coupling $\kappaph/(gLT_0)$.
	The horizontal (finely dashed) line corresponds now to the background noise~$\Po^{JJ}$, which respect to which the noise is normalized.
	}
	\label{fig:FJJ}
\end{figure} 	

In order to show how to detect the behavior of the heat current noise from the temperature fluctuations, we start from Eq.~(\ref{eq:PTT}) together with Eq.~(\ref{app_gencorr}), we use the decomposition introduced in Eqs.~(\ref{eq:Ptotal})-(\ref{eq:Ps}). This procedure leads to
\begin{eqnarray}\label{eq:PTT_ex}
	\frac{\Pp^{TT}(\omega)}{\Po^{JJ}}	
	=
	\frac{1}{\kappa^2} 
		\frac{
	1	}{
	1+(\omega\tauE)^2
	} \left[1
	+ 
	\Fsdir^{JJ}
	+
	\Fsind^{JJ}
\right]	,
\end{eqnarray}
where \mbox{$\Fsdir^{JJ}\equiv\Psdir^{JJ}/\Po^{JJ}$} and \mbox{$\Fsind^{JJ}\equiv\Psind^{JJ}/\Po^{JJ}$} stand for the direct and induced components, respectively, of the source heat-current correlator~\mbox{$\Ps^{JJ}$}, normalized to the equilibrium heat noise
\mbox{$
	\Po^{JJ}
	=
	2D\pi^2
	(k_\mathrm{B}T_0)^3/(3h)
$.}
The explicit expressions for these two terms are found to be
\begin{gather} \label{eq:FJJ_dir_ex}
	\Fsdir^{JJ}
	=
	\frac{3(2-D)}{\pi^2}
	\frac{1}{(2\Omega \sigma )^2} 
	\bigg(\frac{\hbar \Omega}{ k_\text{B} T_0}\bigg)^{\!\!3}
	,
\\ \label{eq:FJJ_ind_ex}
	\Fsind^{JJ}
	=
	\frac{D (\avTp^{\raisebox{-2.5pt}{$\scriptstyle3$}}- T^3_0)}{2T^3_0} 
	+
	\frac{3(1-D)}{\pi^2}\,
	\funH_2 \bigg(\frac{\avTp}{T_0}\bigg)
	.
\end{gather}
Both these terms are shown in Fig.~\ref{fig:FJJ} as a function of the driving frequency for different values of the phonon heat conductance~$\kappaph/(gLT_0)$. Again, it is only the induced part that inherently depends on the probe properties entering via the ratio of thermal conductances~$\kappae/\kappaph$ in~$\avTp$, Eq.~(\ref{eq:ex-T}). While the overall shape of the induced and direct part as a function of the driving frequencies is similar to the one for the electrochemical potential fluctuations, presented in Fig.~\ref{fig:FII}, we note two striking differences.

First and foremost, for the realistic parameters presented here, the induced part~$\Fsind^{JJ}$ of the fluctuations is always smaller than the desired direct part~$\Fsdir^{JJ}$ of the noise. This demonstrates that the proposed injector-detector setup is indeed promising for a readout of heat current correlations via temperature fluctuations. 

Second, it is worth emphasizing that for the heat-current correlator,~\mbox{$\Fs^{JJ}\equiv\Ps^{JJ}/\Po^{JJ}$},  in contrast to the charge-current correlator,~\mbox{$\Fs^{II}\equiv\Ps^{II}/\Po^{II}$}, discussed in Sec.~\ref{sec:mumu_correl_ex}, both contributions~$\Fsdir^{JJ}$ and~$\Fsind^{JJ}$, Eqs.~(\ref{eq:FJJ_dir_ex})-(\ref{eq:FJJ_ind_ex}), are basically insensitive to the transmission probability at the QPC. This can be mostly attributed  to the modified $D$-dependence of the direct part~$\Fsdir^{JJ}$, which, due to its prefactor $(2-D)$, is never strongly suppressed, not even for large QPC-transmissions $D$.  Technically, this stems from the fact  that the heat-current noise, unlike the charge-current noise, contains a so-called \emph{interference} contribution~\cite{Battista2014Aug,Dubois2013Aug} originating from the time-dependent driving. This contribution  has a linear $D$-dependence similar to the purely thermal part with respect to which the expression is normalized here.

The important message of this subsection is that the backaction of the measurement -- resulting in the induced part of the noise --  is not expected to hinder the readout of the heat-current noise, and hence of the temperature fluctuations, under realistic experimental conditions.

\subsubsection{Mixed correlator}

In the discussed setup, the mixed correlators~$\Pp^{\mu T}$ and~$\Pp^{IJ}$ identically vanish. The reason for this is the absence of an electrochemical potential difference, see e.g. Ref.~\cite{Battista2014Dec}. Note that in the detector scheme suggested here, an induced part of this mixed correlator is expected whenever the thermoelectric response-coefficients in Eq.~(\ref{eq:response_coeff}) are non-vanishing.

\subsection{Optimizing parameters}\label{sec:optimization}

In the previous section, we have seen how various tunable parameters -- conductances and capacitances -- influence the probe temperature and the detected fluctuations of temperature and electrochemical potentials in a complex way.  In this section, we give an overview of different aspects that should be taken into account when optimizing these parameters.
The conductance (of importance for the potential fluctuations) is tunable via the QPC transmission $D$.
The situation is more complicated for the thermal conductivities, which depend on temperature themselves. We therefore decide to lead the discussion based on the thermal conductivities $\kappae$ and $\kappaph$ at the base temperature $T_0$. Apart from tuning via the base temperature these two quantities can be tuned by changing the electron-phonon coupling $\Sigma\vol$ via the electron density~\cite{Appleyard1998Oct,Gasparinetti2011May} ($\propto\kappaph$), or by changing the QPC-transmission $D$ ($\propto\kappae$), see Eq.~(\ref{eq:kappa_def}).
These two quantities influence the \textit{magnitude} of the probe temperature and its fluctuations, directly or also indirectly via the probe temperature entering $\kappa$.

The \textit{dynamics} of the fluctuations (important for the bandwidth) depend on the capacitances $C$ and $\CE$ which are tunable through the probe density of states~$\nu$ at the Fermi level. 

\subsubsection{Average quantities}\label{sec:opt_average}
While the average charge current~$\avIs$ is always zero (and so is the average electrochemical potential of the probe,~\mbox{$\avmup=0$}), the average energy current~$\avIs^E$ from the source can be detected via the average probe temperature~$\avTp$, see Eq.~(\ref{eq:tempeq}) and Eq.~(\ref{eq:ex-T}). For this purpose, the difference between the probe temperature with and without the working source should be \textit{maximized} for a fixed average source energy current. 
To meet this goal, the probe thermal conductance~$\kappaph$ due to the coupling to phonons should be small with respect to the electronic thermal conductance $\kappae$. 
%
%

\subsubsection{Magnitude of source fluctuations}

An increase of the probe temperature~$\avTp$, in order to improve its detectability is naturally detrimental for the readout of the charge- and heat-current fluctuations, because it leads to a relative increase of the \textit{induced} noise, $\Psind^{II}$  and $\Psind^{JJ}$. From this point of view, the ratio~$\kappae/\kappaph$ should be \textit{minimized}.
However, also the overall prefactor of the fluctuations depends on this quantity and needs to be optimized. 

In the case of the charge-current fluctuations $\Pp^{II}$, the QPC-transmission~$D$ entering the electronic thermal conductivity $\kappae$ not only needs to be small to ensure that the magnitude of the direct part~$\Psdir^{II}$ is maximized with respect to the induced part~$\Psind^{II}$, but also to increase the overall value of the desired potential fluctuations, see the prefactors in both Eqs. (\ref{eq:Pmumu_ex}) and (\ref{eq:FII_dir_ex}).

\begin{figure}[t!!!]
	\centering
	\includegraphics[width=3in]{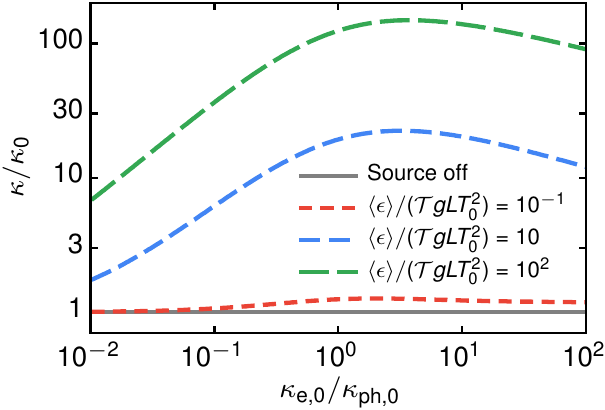}
	\caption{ 
	Thermal conductance at probe temperature~$\kappa$, scaled to its value~$\kappa_0$ at~$T_0$, shown as a function of the ratio of the electron~($\kappae$) and phonon~($\kappaph$) thermal conductances,~$\kappae/\kappaph$, at base temperature $T_0$. We show curves for selected values of~$\langle \epsilon \rangle/(\perT gLT_0^2)$
	-- that is, the average energy of an emitted particle per period,~$\langle \epsilon \rangle/\perT$, related to the electronic heat current response of the probe,~$gL T_0^2$. 
		}
		\label{fig:kappaTp}
\end{figure}

On the other hand, for the detection of the heat-current noise~$\Pp^{JJ}$, the total thermal conductance~$\kappa$ at the probe temperature needs to be small in order to maximize the overall value of $\Pp^{TT}$. 
The interplay of the requirement for a large $\kappa_{\text{ph},0}$ to reduce the induced part of the noise with respect to the direct one, together with the need for a small ~$\kappa$ to maximize the total magnitude of the fluctuations, is crucial from the point of view of the detection-scheme optimization. Thus, in Fig.~\ref{fig:kappaTp} we present the behavior of the thermal conductance at the probe temperature,~$\kappa$,  as a function of the ratio~$\kappae/\kappaph$ for different energies of the emitted particles~$\langle\epsilon\rangle/\perT$ (influencing the probe temperature). The explicit ratio between $\kappa$ and the base-temperature thermal conductance~\mbox{$\kappa_0=\kappae+\kappaph$} can be written as
\begin{equation}\label{eq:kappaTp}
\frac{\kappa}{\kappa_0}
=
\left[
	\dfrac{\kappae}{\kappaph}
	\dfrac{\avTp}{T_0}
	+
	\bigg(\dfrac{\avTp}{T_0}\bigg)^{\!\!4}\right]\left[
	\dfrac{\kappae}{\kappaph}+1\right]^{-1}
,
\end{equation}
using the general, temperature dependent expressions given in Eq.~(\ref{eq:kappa_def}).
It shows  that smaller values of the total thermal conductance~$\kappa$ can be achieved by augmenting the phonon conductance $\kappaph$ with respect to the electronic heat conductance $\kappae$, that is, when \mbox{$\kappae/\kappaph\ll1$}. However, the function saturates, and even decreases again, when $\kappaph$ is small with respect to $\kappae$.

\subsubsection{Bandwidth of noise detection}
For an optimization of the bandwidth, not only the (thermal) conductances, but also the (thermal) capacitances play a role, providing an additional means for the tunability. 

The bandwidth of the charge-current noise detection, improving with increasing \mbox{$1/\tauRC=Dg/C$} [see Eq.~(\ref{eq:Pmumu_ex})], can be optimized by increasing the conductance $\propto D$ or by diminishing the geometric capacitance of the probe, while keeping~$D$ fixed at a value to find the highest achievable magnitude of~$\Psdir^{II}$.

Similarly, a large thermal conductance~$\kappa$, with its dependence on $\kappa_{\text{e},0}$ and  $\kappa_{\text{ph},0}$ at $T_0$ as shown in Fig.~\ref{fig:kappaTp}, is beneficial for the bandwidth of the detection of the heat current noise, $\propto 1/\tauE=\kappa/\CE$. This bandwidth can however also be increased by diminishing the heat capacity~$\CE$, at a fixed  optimized value of $\kappa$.

\section{Conclusions}\label{sec:Conclusions}

In this manuscript we have proposed and analysed an injector-detector setup, suitable for the measurement of charge and heat (or energy) current noise injected by a time-dependently driven electronic current source. For our detection scheme, we  suggest to measure the (macroscopic) temperature and voltage fluctuations arising in a probe due to the operation of the source, in order to extract the charge and energy current fluctuations from them. Using a Boltzmann-Langevin approach together with an appropriate correlator decomposition scheme, we show how the frequency-dependent fluctuations of the macroscopic probe properties are related to the current correlators to be detected. We carefully analyze the backaction effects of the detection scheme on the signal to be extracted. 

As a proof-of-principle example, we investigate as the current source to be analyzed an ideally time-dependently driven mesoscopic capacitor emitting a sequence of well-separated quantized charge pulses. Based on this example, we show the behavior of the different contributions to the fluctuations as a function of the setup parameters and thereby indicate regimes in which a detection of heat current fluctuations appears feasible. Thr experimental detection of recent theoretically investigated energetic properties of electronic single-particle sources~\cite{Moskalets2014May,Battista2014Aug,Battista2014Dec,Rossello2015Mar} and (mixed) heat-current fluctuations more generally~\cite{Crepieux2014Dec} is hence made tangible by our proposed setup.

\acknowledgments
We thank Christian Glattli for valuable comments on energetic properties of single-electron pulses and Gwendal F\`eve  and Fran\c{c}ois Parmentier for useful discussions about the experimental feasibility of this setup.
Funding from the Knut and Alice Wallenberg Foundation through the Academy Fellow program (J.S., N.D. and M.M.) and from the Swedish VR is gratefully acknowledged. M.\,M. also acknowledges financial support from the Polish Ministry of Science and Education through a young scientist fellowship (0066/E-336/9/2014).

\appendix

\section{\label{app:transport_theory}Time-dependent transport theory}

To calculate charge and energy currents, as well as respective noise correlators in a time-periodically driven mesoscopic conductor, we use a scattering matrix approach, which we briefly introduce here. 
In general, the properties of a time-dependently driven mesoscopic conductor are described by the Floquet scattering matrix~\cite{Moskalets2011Sep}. In the considered system, that is, for a two-terminal conductor supporting a single channel, the elements of the scattering matrix are $\SF_{\alpha\beta}(E_n,E_m)$. They represent the amplitude to scatter from energy $E_m$ in terminal $\beta$ to energy $E_n$ in terminal $\alpha$, with \mbox{$E_n=E+n\hbar \Omega$} and \mbox{$\alpha,\beta=\text{L},\text{p}$}. 
As a result, the annihilation operators~$\op{b}_\alpha(E)$ for outgoing particles at energy $E$ in terminal $\alpha$ can be related to the annihilation operators~$\op{a}_\beta(E_n)$ for incoming particles at energy $E_n$ in terminal $\beta$ by means of the scattering matrix as follows
\begin{equation}
	\op{b}_{\alpha}(E)
	= 
	\sum\limits_\beta  
	\sum\limits^{\infty}_{n= - \infty} 
	\!
	\SF_{\alpha \beta}(E,E_{n})  
	\op{a}_{\beta}(E_{n}).
	\label{oprel}
\end{equation}
The operators for the \emph{charge}, $\opI_{\alpha}(t)$, and \emph{energy}, $\opI^E_{\alpha}(t)$, currents in terminal~$\alpha$  can be written in terms of the corresponding annihilation and  creation operators as
\begin{multline}\label{App_current}
	\opI_{\alpha} (t) 
	=
	-
	\frac{e}{h} 
	\iint\! \intd E \intd E^\prime \, 
	\text{e}^{i (E-E^\prime)t/\hbar}
\\
	\times 
	\Big\{ 
	\op{b}_{\alpha}^{\dagger}(E)\, \op{b}_{\alpha} (E^\prime) 
	- 
	\op{a}_{\alpha}^{\dagger}(E)\, \op{a}_{\alpha} (E^\prime) 
	\Big\}
	,
\end{multline}
and
\begin{multline}\label{App_Ecurrent}
	\opI^{E }_{\alpha} (t) 
	=
	\frac{1}{h} \iint\! \intd E \intd E^\prime \, 
	\dfrac{(E+E^\prime)}{2} 
	\text{e}^{i (E-E^\prime)t/\hbar}
\\ 
	\times 
	\Big\{ 
	\op{b}_{\alpha}^{\dagger}(E)\, \op{b}_{\alpha} (E^\prime) 
	- 
	\op{a}_{\alpha}^{\dagger}(E)\, \op{a}_{\alpha} (E^\prime) 
	\Big\}
	.
\end{multline}
The \emph{heat} current operator can then be obtained from 
\begin{equation}
	\hat J_{\alpha}
	=
	\opI^E_{\alpha}
	+
	\overline{\mu}_{\alpha}\opI_{\alpha}/e
\end{equation}
Employing the current operators in Eqs.~(\ref{App_current})-(\ref{App_Ecurrent}) together with the operator relation in Eq.~(\ref{oprel}), the charge and energy currents averaged over one period of the driving are derived along standard lines, see e.g. Ref.~\cite{Moskalets2011Sep}, 
\begin{gather}\label{Iav}
	\avI_{\alpha}
	=
	-
	\frac{e}{h}
	\int\! \intd E\,
	\Big[
	\widetilde{f}_{\alpha}(E)
	- 
	f_{\alpha}(E)
	\Big]
	,
\\\label{IEav}
	\avI^E_{\alpha}
	=
	\frac{1}{h}
	\int\! \intd E E
	\Big[
	\widetilde{f}_{\alpha}(E)
	- 
	f_{\alpha}(E)
	\Big]
	,
\end{gather}
with 
\mbox{$
	f_\alpha(E)
	=
	\big\{1+\text{exp}\big[(E-\avmu_\alpha)/(\kB T_\alpha)\big]\big\}^{-1}
	$}
denoting the Fermi distribution in the terminal~$\alpha$, and
\begin{equation}
	\widetilde{f}_{\alpha}(E)
	=
	\sum\limits_{\beta}  
	\sum\limits^{\infty}_{n= - \infty}  
	\!
	\big|\SF_{\alpha\beta}(E,E_n)\big|^2 f_{\beta}(E_n) 
	.
\end{equation}

On the other hand, the low-frequency, period-averaged correlations between currents of type~\mbox{$A,B=I,J$} in terminal $\alpha$ are defined as
\begin{equation}\label{barecorr}
	\mathcal{P}^{AB}_{\alpha}
	= 
	\frac{1}{2}
	\int\limits^{\perT}_{0}\!\frac{\intd t}{\perT} 
	\int\limits^{\infty}_{-\infty}\!\!\intd t^\prime\,
	\big\langle 
	\big\{\delta \hat{A}_\alpha(t), \delta \hat{B}_\alpha(t+t^\prime)\big\} 
	\big\rangle, 
\end{equation}
where \mbox{$\delta \hat{A}_{\alpha}(t)\equiv\hat{A}_{\alpha}(t)-\overline{A}_{\alpha}$}. Following the standard procedure outlined, e.g., in Refs.~\cite{Moskalets2011Sep,Battista2014Aug,Moskalets2014May}, we derive the~expression for charge, energy and mixed current auto-correlations, conveniently written in a general form as, 
\begin{widetext}
\begin{align}\label{corrFloq}
	\mathcal{P}_{\alpha}^{AB}
	= 
	\frac{1}{h} 
	\int\! \intd E\,  
	\bigg\{
	&
	x_Ax_B f_\alpha(E)\big[1-f_\alpha(E)\big]
	-
	\sum\limits_{n=-\infty}^\infty 
	\!
	\big(x_Ax_{B,n}+x_{A,n}x_B\big)
	\big|\SF_{\alpha\alpha}(E,E_n)\big|^2 
	f_{\alpha}(E_n)\big[1-f_\alpha(E_n)\big]
\nonumber\\	
	+&
	\sum\limits_{n=-\infty}^\infty 
	\!
	\frac{x_Ax_{B,n}+x_{A,n}x_B}{2} 
	\sum\limits_{\beta\gamma}
	\sum\limits_{k,q=-\infty}^\infty 
	\!
	\SFcon_{\alpha\beta}(E,E_k)
	\SF_{\alpha\beta}(E_n,E_k)
	\SF_{\alpha\gamma}(E,E_q)
	\SFcon_{\alpha\gamma}(E_n,E_q)
\nonumber \\
	&
	\hspace*{278pt}
	\times
	f_{\beta}(E_k)
	\big[1-f_{\gamma}(E_q)\big]
	\bigg\}
	,
\end{align}
where we have introduced a shorthand notation: $x_{I,n}= x_{I}\equiv -e$ and  \mbox{$x_{J,n}\equiv E_n-\avmup$}  (with \mbox{$x_J\equiv x_{J,0}$}).

\section{Mesoscopic capacitor -- induced correlations}\label{app:meso_induced}

\subsection{Chemical potential correlators}

To obtain the induced part~$\Psind^{II}$, we start from the full expression for the charge-current correlator~(\ref{app_gencorr}), and we omit the direct part~$\Psdir^{II}$. The remaining correlator~\mbox{$\Fsind^{II}\equiv\Psind^{II}/\Po^{II}$} (for convenience divided by the equilibrium thermal noise of the QPC,~\mbox{$\Po^{II}=2Dgk_\text{B}T_0$})  can then be written as
\begin{align} 
	\Fsind^{II}
	=\ &
	\frac{1-D}{2\kB  T_0}
	\int\! \intd E
	\sum\limits_{n=-\infty}^\infty 
	\!
	\big|S_{n}\big|^2\, 
	\big[\fp(E)-\fL(E)\big] 
	\big[1-2\fL(E_n)\big] 
\nonumber \\
	&+
	\frac{D}{2\kB T_0}
	\int\! \intd E\,
	\Big\{
	\fp(E)[1-\fp(E)]
	-
	\fL(E)[1-\fL(E)]
	\Big\}
	.
\end{align}
\end{widetext}
Performing the last integral, and inserting the explicit expression for the frozen scattering matrix~$S_n$, Eq.~(\ref{eq:scattering}), and we find 
\begin{align}\label{eq:FsII_ex}
	\Fsind^{II}
	=\ & 
	-\frac{1-D}{\kB T_0} 
	(2\Omega  \sigma )^2  
	\int\! \intd E\, 
	\big[ \fp(E)-\fL(E)\big] 
\nonumber\\
	&\hspace*{22pt}
	\times
	\sum\limits_{n>0} 
	\text{e}^{-2n\Omega \sigma} 
	\big[\fL(E_n)+\fL(E_{-n})\big]
\nonumber \\
	&
	+ \frac{D (\avTp- T_0)}{2T_0}
	+(1-D)\,\funH_0\bigg(\frac{\avTp}{T_0}\bigg) 
	,
\end{align}
Making use of the adiabatic and low-temperature conditions given by Eq.~(\ref{eq:ad_cond}), the first term of the equation above can be expanded up to leading order in~\mbox{$\Omega\sigma\ll1$}. One finds then that this term is negligibly small compared to the remaining two terms in the last line of Eq.~(\ref{eq:FsII_ex}). With this, Eq.~(\ref{eq:FII_ind_ex}) from the main text is obtained.

\subsection{Temperature correlators}

The induced part~$\Psind^{JJ}$ of the heat-current correlator~$\Pp^{JJ}$ can be derived in an analogous manner as discussed in the section above.
Again, starting from the full expression for the correlator~(\ref{app_gencorr}), the normalized induced component~\mbox{$\Fsind^{JJ}\equiv\Psind^{JJ}/\Po^{JJ}$} (with 
\mbox{$
	\Po^{JJ}
	=
	2D\pi^2
	(k_\mathrm{B}T_0)^3/(3h)
$}
representing the equilibrium heat noise of the QPC) can be written as
\begin{widetext}
\begin{align} 
	{\mathcal F}^{JJ}_{\text{s,ind}} 
	=\ &
	\frac{3(1-D)}{2\pi^2(\kB T_0)^3}
	\int\! \intd E\, E^2\!
	\sum\limits_{n=-\infty}^\infty 
	\!
	\big|S_{n}\big|^2\, 
	\big[\fp(E)-\fL(E)\big] 
	\big[1-2\fL(E_n)\big]
\nonumber \\
	&
	+
	\frac{3D}{2\pi^2(\kB T_0)^3}
	\int\! \intd E\, E^2
	\Big\{
	\fp(E)[1-\fp(E)]
	-\fL(E)[1-\fL(E)]
	\Big\}
	.
\end{align}
Performing integrals and rearranging terms similarly as in Eq.~(\ref{eq:FsII_ex}), we get 
\begin{align} 
	{\mathcal F}^{JJ}_{\text{s,ind}} 
	=\ & 
	-\frac{3(1-D)}{\pi^2(k_\text B T_0)^3} 
	\int\! \intd E\, E^2\,
	\big[ \fp(E)-\fL(E)\big] 
	\sum\limits_{n>0} 
	\text{e}^{-2n\Omega \sigma} 
	\big[\fL(E_n)+\fL(E_{-n})\big]
	\nonumber \\
	&
	+
	\frac{D (\avTp^{\raisebox{-2.5pt}{$\scriptstyle3$}}- T^3_0)}{2T^3_0} 
	+
	\frac{3(1-D)}{\pi^2}\,
	\funH_2 \bigg(\frac{\avTp}{T_0}\bigg)
	.
\end{align}
Finally, using the low-temperature adiabatic expansion of the first term in the equation above, and consequently retaining only the terms in the second line, we get Eq.~(\ref{eq:FJJ_ind_ex}) in the main text.
\end{widetext}


%

\end{document}